\documentclass[double]{article}
\usepackage{times}
\usepackage{algorithm}
\usepackage{algorithmic}
\usepackage{wrapfig}
\usepackage{verbatim}
\usepackage{amsmath}
\usepackage{graphicx}
\usepackage{subcaption}
\usepackage[active]{srcltx}
\usepackage{color}
\usepackage{lineno}
\usepackage{dirtytalk}
\usepackage{amsthm}
\usepackage{authblk}
\usepackage{listings}
\usepackage{tikz}
\usepackage{longtable}
\usepackage{setspace}
\usepackage{booktabs}
\usepackage{multirow}
\usepackage{siunitx}
\usepackage{rotating}
\usepackage{adjustbox}
\usepackage{pgfplots}
\usepackage{threeparttable}
\usepackage{caption}
\usepackage{array}
\usepackage{textcomp}
\usepackage{dirtytalk}
\usetikzlibrary{shapes.geometric, arrows, positioning}
\pgfplotsset{compat=1.18}

\usepackage{epsfig,graphicx,amsfonts}
\usepackage{amsmath,amsthm,amssymb}
\usepackage{xcolor}
\usepackage{adjustbox}
\usepackage{multirow}
\usepackage{setspace}
\usepackage{hyperref}
\usepackage{comment}

\newcommand{\review}[1]{{#1}}

\long\def\comment #1\commentend{}

\begin{document}

\title{\Large Bi-National Academic Funding and Collaboration Dynamics:
The Case of the German-Israeli Foundation
}

\author{
Amit Bengiat$^{1}$,
Teddy Lazebnik$^{2,3}$,
Philipp Mayr$^{4}$,
Ariel Rosenfeld$^{1,*}$ \\
$^{1}$ Department of Information Science, Bar-Ilan University, Israel \\
$^{2}$ Department of Information Systems, University of Haifa, Israel \\
$^{3}$ Department of Computing, J\"{o}nk\"{o}ping University, Sweden \\
$^{4}$ Knowledge Technologies for the Social Sciences,
GESIS -- Leibniz-Institute for the Social Sciences \\
$^{*}$ Corresponding author: \texttt{ariel.rosenfeld@biu.ac.il}
}

\date{ }
\maketitle

\begin{abstract}
\noindent
Academic grant programs are widely used to motivate international
research collaboration and boost scientific impact across borders.
Among these, bi-national funding schemes -- pairing researchers from
just two designated countries -- are common yet understudied compared
with national and multinational funding. In this study, \review{we explore whether bi-national programs are associated with new collaborations and lasting partnerships.} To this end, we
conducted an observational bibliometric case study of the German--Israeli Foundation
(GIF), covering 642 grants, 2,386 researchers, and 52,847 publications.
Our results show that GIF funding is associated with increased
co-authorship behavior during, and even slightly before, the grant period, but
it is rarely linked with long-lasting co-authorship patterns
that persist once the funding concludes. By tracing co-authorship
before, during, and after the funding period, clustering collaboration
trajectories with temporally-aware K-means, and predicting cluster
membership with ML models, we find that 45\% of teams
with no prior co-authored publications become active while
funded, yet activity declines rapidly post-award; roughly
one-third sustain longer-term co-authorship activity, whereas only a
small subset sustains high, lasting co-authorship activity. Moreover, the team's composition and bibliometrics are found to be largely insufficient in predicting co-authorship activity during and after the grant. \\ 

\noindent
\textbf{Keywords}: international collaboration; funding; collaboration
sustainability; scientometrics; bi-national funding.
\end{abstract}

\maketitle \thispagestyle{empty}
\pagestyle{myheadings} \markboth{Accepted to \emph{Scientometrics} as of July 24$^{th}$, 2026.}{Accepted to Scientometrics as of July 24$^{th}$, 2026.}
\setcounter{page}{1}
\onehalfspacing

\section{Introduction}
\label{sec:intro}

Academic grant funding is often considered to stimulate and reinforce international collaborations, with prior work reporting substantial effects on academic
collaboration networks and research impact
\cite{clark2012investments,Leydesdorff2017The,zhou2020depth}.
Prior literature in this field predominantly explored large-scale,
multi-national funding mechanisms such as the European Union Framework
Programmes (e.g., Horizon
2020)\footnote{\url{https://research-and-innovation.ec.europa.eu/funding/funding-opportunities/funding-programmes-and-open-calls/horizon-europe_en}},
which are explicitly directed at large-scale collaboration across many
countries and institutions, and sometimes even mandate it, demonstrated
their overall effects \cite{hoekman2013acquisition,Morillo2019Collaboration}.
These prior studies largely credit multi-national funding programs with
enhanced international co-authorship, network diversity, and research
impact, especially for smaller or less-connected countries \cite{vieira2023influence, adams2024national}. 


While multi-national funding is the most studied mechanism for
fostering international collaboration, other funding schemes may
promote international collaborations as well. 
National funding, for example, has been studied in a few prior works as a potential catalyst
for international collaborations \cite{Wang2024Elevating,Abramo2022The}.
However, as these programs target excellence within a country's own
scientific system, such as the individual grants program of the Israeli
Science Foundation\footnote{\url{https://www.isf.org.il/}} or the
German Research Foundation (Deutsche Forschungsgemeinschaft,
DFG)\footnote{\url{https://www.dfg.de/en}}, their ability to foster
international partnerships seems significantly more modest compared to
multi-national funding schemes \cite{jacob2011impact,c2,c3}. Interestingly, despite their global
prevalence, bi-national funding programs remain
under-researched compared to both national and multi-national schemes.

Bi-national funding schemes operate under very different dynamics than
national or multinational ones. In particular, international
collaborations are explicitly targeted through bi-national funding (as
opposed to national schemes), yet these are mostly dyadic rather than
consortium-based (as opposed to multi-national schemes), presumably
leading to different collaboration dynamics and outputs.
For example, evidence from the Indo-German case
\cite{mir2024,indo-german_2024} suggests that such schemes do generate
measurable bi-national co-authorship activity, yet it remains largely unclear \review{whether they are associated with genuinely new connections or primarily with the reinforcement of existing ones}, and
to what extent co-authorship persists beyond the funded period.
Considering the former, existing literature suggests that research funding typically
consolidates existing communities rather than bridges disconnected ones \cite{davies2022research,Hicks2019Network,defazio2009funding}. However, these dynamics are primarily observed within large-scale and national funding programmes, whether they apply to the smaller, dyadic structure of bi-national schemes remains an open question.
Considering the latter, Network Theory suggests that resilient
scientific ties are built through repeated, multi-contextual
interactions that gradually accumulate shared infrastructure, mutual
dependency, and interpersonal trust \cite{c5}. Short-term,
project-based funding, however, is often assumed to generate \say{weak ties} - connections activated by an external stimulus but lacking the structural embeddedness needed for long-term continuity
\cite{sydow2018projects,meirmans2024competition,granovetter1973strength}. Consistent with this view, several
empirical studies have demonstrated that grant-induced collaboration
tends to peak during the funding window and decline appreciably
thereafter \cite{kosmützky2021varieties,urbanovics2024path}. However, because this decay is predominantly observed within multi-national and national funding programmes, \review{whether, under bi-national funding, these transient ties persist as enduring partnerships remains an open question.} 

\review{Our study seeks to address two open questions: 1)~Are bi-national grants associated with the formation of new co-authorship ties between the involved countries, or primarily with the reinforcement of ties between already-connected groups? and 2)~Is bi-national co-authorship activity elevated before and during the grant period, and does it persist once the funding concludes?}
To address these questions, we conduct a comprehensive bibliometric
case study of the highly successful and long-established German-Israeli
Foundation for Scientific Research and Development (often abbreviated
simply as \textit{GIF})
program\footnote{\url{https://www.gif.org.il/}}. The GIF program was
established in 1986 through an agreement between the Ministries of
Science of the Federal Republic of Germany and the State of Israel, as
an important instrument complementing the wider, fruitful
socio-economic ties between the two countries
\cite{gif_about,yair2020hierarchy,yair2019culture,yair2023emotional}.
GIF's mandate is to strategically fund civilian research and
development across both basic and applied research, maintaining
disciplinary neutrality by evaluating proposals from all scientific
fields without quotas or thematic restrictions
\cite{jewishvirtuallibrary}. Since its inception, the foundation has
supported over 2,000 research projects with total funding surpassing
\texteuro270 million, involving approximately 4,000 scientists from
300 academic institutions \cite{gif_about}.
Aligned with the common practice in prior bibliometric studies of research collaboration \cite{subramanyan1983bibliometric}, we use co-authorship as a proxy for deeper collaborative
relationship such as joint intellectual effort, shared resources, and mutual commitment
\cite{kahn2018co}. Nevertheless, other collaborative activities, such as student
exchanges, shared datasets, joint infrastructure, or follow-on grant
applications are not reflected through this proxy and hence considered outside the scope of this work \cite{bozeman2004scientists,sonnenwald2007scientific}.

Overall, the present study makes three primary contributions. First, it provides the most comprehensive bibliometric study of a bi-national funding program to date -- a funding scheme that, despite its
global prevalence, has received little empirical attention thus far. Second,
it presents a nuanced temporal perspective on co-authorship dynamics by tracing co-authorship activity throughout the full grant lifecycle, transcending the typical reliance on one-shot measures or aggregated pre- and post-funding comparisons
\cite{jacob2011impact,c2}. Third, by integrating temporally-aware clustering with predictive modeling, the study demonstrates that standard team composition and scientometrics are largely insufficient in predicting long-term co-authorship dynamics and survival.

The remainder of this study is organized as follows.
Section~\ref{sec:methods} formally outlines the data collection and
analytical approach. Section~\ref{sec:results} presents the obtained
results in terms of descriptive statistics, clustering analysis, and
ML-based prediction for cluster association. Finally,
Section~\ref{sec:discussion} discusses the policy implications of the
results and suggests avenues for future work.

\section{Methods and Materials}
\label{sec:methods}

In this section, we formally introduce the data collection process and
analysis approach employed.

\subsection{Data Collection}
Data collection took place for two main sources: grant extraction and researcher identification, as well as bibliometric and demographic data of those researchers.

\subsubsection{Grant extraction and researcher identification}

Our database draws information from two sources: the official GIF
website and OpenAlex, a bibliographic catalogue of scientific papers,
authors, and institutions with over 250 million scholarly works from
250 thousand sources \cite{priem2022openalex,culbert2025reference}.
Initially, we conducted a manual extraction of the complete grant
portfolio from the GIF official website, capturing 1,193 unique grant
records across all available funding cycles and research programs.
From these 1,193 grant records, only 647 (54.2\%) are explicitly
associated with a German-Israeli \say{team}. Namely, each of these
grants was awarded to at least a single Israeli grant recipient and at
least a single German recipient, as indicated by their listed
institutional affiliations. These grants alone are considered in our
subsequent analysis. For each grant, we extracted its key available
features as summarized in Table~\ref{table:grant_info}.

Next, we cross-referenced each researcher with their bibliometric
data available through the OpenAlex scholarly database API
\cite{culbert2025reference}. This process employed a three-step
researcher identification algorithm designed to handle the challenges
of international name variations and institutional affiliations in our
setting: For each researcher's name, we first implemented exact
institutional matching using an expert manually-curated dictionary of
over 50 institutional name variants, accounting for the different
languages (German, Hebrew, English), common abbreviations, and
historical name changes. The dictionary included mappings for major
German universities (e.g., \say{Johann Wolfgang Goethe University}
was mapped to common variants such as \say{Goethe-Universit\"{a}t
Frankfurt} and \say{University of Frankfurt}) and Israeli institutions
(e.g., \say{Hebrew University} was mapped to \say{Hebrew University
of Jerusalem}). Only when an exact match of both the researcher's name and
institution was not found, a second step was performed bidirectional
word-subset matching with text normalization, including accounting for
diacritical marks (\"{a}, \"{o}, \"{u} to ae, oe, ue),
standardization of transliterations, handling of common academic title
variations, and systematic accounting of known institutional acronyms
(for example, HUJI, BGU, BIU, TAU). If the above results in no match,
a final step was conducted using historical affiliation verification by
analyzing researchers' complete OpenAlex institutional histories. This
step intends to resolve cases where current affiliations differed from
those recorded by the GIF. Only in rare cases (about 1\%), where this
process failed to produce an exact match, a manual annotation was
performed by the first author, who accessed the institutional faculty
directories and records online. This approach resulted in 99.7\%
researcher matching, leading to 8 researchers being removed from the
dataset. Grants associated with these eight researchers were retained in the dataset, as they remained part of an explicit German-Israeli team following the researchers' exclusion.To assess the reliability of the matching procedure, we perform a simple error analysis. First, a manual examination of the unmatched 8 researchers suggests that they are associated with either poor biographical information in OpenAlex or have very common names without clearly distinguishing institutional markers. Second, a random sample of 100 automatically-matched researchers was manually examined, comparing their matched entry to their online institutional faculty records, revealing no discrepancies.

Figure~\ref{fig:matching_flowchart} illustrates the researcher matching
procedure.

\begin{figure}[H]
\centering
\begin{tikzpicture}[
    node distance=1.8cm,
    box/.style={rectangle, rounded corners, draw=black, thick,
                text width=7cm, align=center, minimum height=1cm,
                fill=gray!10},
    arrow/.style={->, thick, >=stealth},
    side/.style={rectangle, draw=black, dashed, text width=4.5cm,
                 align=center, minimum height=1cm, fill=white,
                 font=\small}
]
\node[box] (A) {Researcher name and institution extracted from GIF
                website};
\node[box] (B) [below=1.8cm of A]
  {\textbf{Step 1:} Exact institutional matching \\
   (curated dictionary of 50+ name variants across German, Hebrew, and
   English)};
\node[box] (C) [below=1.8cm of B]
  {\textbf{Step 2:} Bidirectional word-subset matching \\
   (diacritics normalisation, transliterations, acronyms, title
   variations)};
\node[box] (D) [below=1.8cm of C]
  {\textbf{Step 3:} Historical affiliation verification \\
   (complete OpenAlex institutional histories)};
\node[box] (E) [below=1.8cm of D]
  {\textbf{Manual annotation} ($\sim$1\% of cases) \\
   (faculty directories, online records)};
\node[box, fill=green!15] (F) [below=1.8cm of E]
  {Researcher matched\\
   \textit{(2,386 researchers)}};
\draw[arrow] (A) -- (B);
\draw[arrow] (B) -- (C);
\draw[arrow] (C) -- (D);
\draw[arrow] (D) -- (E);
\draw[arrow] (E) -- (F);
\node[side] (s1) [right=1.5cm of B]
  {Matched via institutional\\
   record (single entry)\\
   $\rightarrow$ \textbf{116 matched}};
\node[side] (s2) [right=1.5cm of C]
  {Unique institutional match found\\
   $\rightarrow$ \textbf{1,673 matched}};
\node[side] (s3) [right=1.5cm of D]
  {Multiple records combined\\
   across institutional histories\\
   $\rightarrow$ \textbf{530 matched}};
\node[side] (s4) [right=1.5cm of E]
  {OpenAlex record located via\\
   external sources\\
   $\rightarrow$ \textbf{67 matched}};
\draw[->, dashed] (B.east) -- (s1.west);
\draw[->, dashed] (C.east) -- (s2.west);
\draw[->, dashed] (D.east) -- (s3.west);
\draw[->, dashed] (E.east) -- (s4.west);
\end{tikzpicture}
\caption{Researcher identification and matching procedure.}
\label{fig:matching_flowchart}
\end{figure}

\subsubsection{Bibliometric and demographic data}

As part of the researcher profile cross-referencing process, all
indexed peer-reviewed publications and standard bibliometrics, as
detailed in Table~\ref{table:grant_info}, were retrieved through
OpenAlex's API. Finally, as the researchers' gender is not explicitly
mentioned in our two data sources, we adopted the name-based
gender-identification model of \cite{hu2021name}, allocating gender
only when the model predicted the gender with 95\% confidence or more.
\review{We used this model to align with previous scientometric studies also
using this model for gender identification
\cite{alexi2024scientometrics,lazebnik2025lonely,rosenfled2025academic}.
Of the 2,386 researchers in the dataset, gender was
successfully classified for 1,716 researchers (71.9\%), comprising
1,341 males (56.2\%) and 375 females (15.7\%). The remaining 670
researchers (28.1\%) were not classified and were assigned an
indeterminate gender label (N). Unclassified cases arose from two
sources: first names absent from the gender database (336 cases,
50.1\%) and cases where the model's confidence fell below the 95\%
threshold (334 cases, 49.9\%).} Accordingly, gender is treated as a supplementary
descriptive variable in our analysis rather than a core predictor,
and all gender-related findings should be interpreted with appropriate
caution.

Further examination of the 647 grants and their research teams led to
5 of them being classified as outliers in terms of their publication
volume, which exceeded the average by a factor of at least 9 
(range: 9.4--66.8), and subsequently removed from the data. Overall, our dataset comprises 642 grants, 2,386 researchers, and 52,847 unique
publication records spanning the years 1986--2024, with less than
0.4\% missing values.

Figure~\ref{fig:flowchart} illustrates the grant extraction and filtering process.

\begin{figure}[!ht]
\centering
\begin{tikzpicture}[
    node distance=1.8cm,
    box/.style={rectangle, rounded corners, draw=black, thick,
                text width=7cm, align=center, minimum height=1cm,
                fill=gray!10},
    arrow/.style={->, thick, >=stealth},
    side/.style={rectangle, draw=black, dashed, text width=4.5cm,
                 align=center, minimum height=1cm, fill=white,
                 font=\small}
]
\node[box] (A) {1,193 grant records extracted from GIF website};
\node[box] (B) [below=1.8cm of A]
  {\textbf{Step 1:} Nationality filtering \\
   (retain only grants with at least one Israeli and one German recipient)};
\node[box] (C) [below=1.8cm of B]
  {\textbf{Step 2:} Outlier removal \\
   (exclude grants with anomalous publication volume)};
\node[box, fill=green!15] (D) [below=1.8cm of C]
  {642 grants};
\draw[arrow] (A) -- (B);
\draw[arrow] (B) -- (C);
\draw[arrow] (C) -- (D);
\node[side] (s1) [right=1.5cm of B]
  {Excluded: 546 grants (45.8\%) $\rightarrow$ 647 grants retained};
\node[side] (s2) [right=1.5cm of C]
  {Excluded: 5 grants $\rightarrow$ 642 grants retained};
\draw[->, dashed] (B.east) -- (s1.west);
\draw[->, dashed] (C.east) -- (s2.west);
\end{tikzpicture}
\caption{Grant extraction and filtering process.}
\label{fig:flowchart}
\end{figure}

\begin{table}[!ht]
\centering
\begin{threeparttable}
\begin{tabular}{p{4cm}p{9cm}}
\toprule
\textbf{Feature Category} & \textbf{Variables} \\
\midrule
\textbf{Grant Information} & \\
\addlinespace[0.5ex]
Title & Project title as extracted from GIF website \\
\addlinespace
Researchers & Principal investigator names (used for OpenAlex
matching) \\
\addlinespace
Program & Research program classification \\
\addlinespace
Research Area & Scientific field or domain of the research project \\
\addlinespace
Proposal Year & Year the grant proposal was submitted and approved \\
\addlinespace
Institution Country & Country classification (Germany/Israel) \\
\addlinespace[1ex]
\textbf{OpenAlex Information} & \\
\addlinespace[0.5ex]
Researcher Name & OpenAlex display name (may differ from original
name) \\
\addlinespace
Original Name & Name from GIF database (used for matching) \\
\addlinespace
Institution & Home institution from GIF (used for institutional
matching) \\
\addlinespace
OpenAlex ID & Author identifier (e.g., A5080334402) \\
\addlinespace
ORCID & Open Researcher and Contributor ID when available \\
\addlinespace[1ex]
\textbf{Bibliometric Metrics} & \\
\addlinespace[0.5ex]
H-Index & Citation-based impact measure \\
\addlinespace
i10-Index & Number of publications with at least 10 citations \\
\addlinespace
Total Papers & Complete publication count \\
\addlinespace
Total Citations & Aggregate citation count \\
\addlinespace[1ex]
\textbf{Institutional Affiliations} & \\
\addlinespace[0.5ex]
Institution Name & Affiliated institution names \\
\addlinespace
Institution Country & Institution country codes \\
\addlinespace[1ex]
\textbf{Publication Details} & \\
\addlinespace[0.5ex]
Paper Title & Individual publication titles \\
\addlinespace
Publication Year & Year of publication \\
\addlinespace
Journal Information & Journal name and ISSN \\
\addlinespace
DOI & Digital Object Identifier \\
\addlinespace
Citation Count & Per-paper citation metrics \\
\addlinespace
Publication Type & Article, review, book chapter, etc. \\
\addlinespace
Open Access & Open access availability status \\
\addlinespace[1ex]
\textbf{Collaboration Data} & \\
\addlinespace[0.5ex]
Collaborator Names & Co-author names separated by vertical bars \\
\addlinespace
Collaborator IDs & OpenAlex IDs of co-authors (normalised to primary
IDs) \\
\addlinespace[1ex]
\textbf{Gender Information} & \\
\addlinespace[0.5ex]
First Name & Extracted first name for gender prediction \\
\addlinespace
Predicted Gender & Predicted gender (M/F/N) \\
\addlinespace
Gender Confidence & Confidence score and percentage \\
\bottomrule
\end{tabular}
\end{threeparttable}
\caption{Features of the grant, researcher, and publication in the
dataset.}
\label{table:grant_info}
\end{table}

\subsection{Analytical Approach}

Since our study focuses on grants with an explicit German-Israeli team
of recipients, we consider a publication to be \say{co-authored by the
team} if and only if at least one of the Israeli grant recipients and
at least one of the German grant recipients are listed in its byline.
As such, our subsequent analysis explores the dynamics of these
co-authored publications.

We start our analysis with a descriptive analysis of the awarded grants,
focusing on their temporal co-authorship dynamics (i.e., the
rate of grants with at least one co-authored publication for a given
year), number of team members per awarded grant, and the distribution
of key bibliometrics across funded teams. Then, we temporally divided the data into three distinct periods (pre-grant, during-grant, and post-grant) and performed a trend and survival analysis to test for temporal trends.
Then, we conduct a
multi-time clustering analysis and implement three distinct ML
approaches to predict cluster membership based on their researchers'
characteristics. Clustering is used as a data-driven instrument to
uncover prototypical collaboration dynamics of German-Israeli teams
within the GIF's funding program. The ML-based cluster membership
prediction is used as a complementary tool to explore which teams are a prior more likely to present different collaboration dynamics.

\subsubsection{Trend and Survival Analysis}

We first temporally divided the data into
three distinct periods: \textit{Pre-grant period} -- ten years before
the grant was awarded (denoted as $\ldots, -2, -1$);
\textit{During-grant period} -- the years during which the grant is
active (typically 3 years, denoted $0, 1, \ldots$) extended by 2
years to allow sufficient time for outcomes to be written, reviewed,
and published (aligned with the common practice in prior work \cite{lyman2013three,sarigol2017quantifying}); and
\textit{Post-grant period} -- ten years from the conclusion of the
grant period onward. 

To test for temporal trends in co-authorship rates, we applied Ordinary Least Squares linear regression \cite{fisher1992statistical} separately to the pre-grant decade ($t\in[-10,-1]$) and the post-grant decade ($t\in[6,16]$), and compared the resulting slopes using a z-test. To assess co-authorship durability following the grant period, we conducted a Kaplan-Meier survival analysis
\cite{kaplan1958nonparametric} among the teams that were active
during the grant period, treating the end of co-authorship activity
as the event of interest.
\review{Specifically, survival time was measured in years from the end of the during-grant window. For each team, the survival time was defined as the number of years from grant end to the team's final post-grant co-authored publication, and the event of interest (cessation of co-authorship) was treated as observed at that year; teams with no post-grant co-authored publication were assigned a survival time of zero. Teams whose final co-authored publication occurred in the last year of the post-grant observation window were right-censored, as continued co-authorship beyond the window could not be ruled out.}

\subsubsection{Temporal clustering and statistical testing}

Considering the temporal division outlined above, the clustering analysis was employed using the Dynamic Time Warping (DTW) metric \cite{sakoe1978dynamic} combined
with the K-means algorithm \cite{macqueen1967some}, utilising
grid-search hyperparameter optimisation using the silhouette score
metric \cite{rousseeuw1987silhouettes} to find the optimal value for
the number of clusters, $K$.

For robustness, we replicate the above analysis by varying both the distance metric and the temporal window. Specifically, we replaced DTW with the discrete Fr\'{e}chet distance
\cite{eiter1994computing,alt1995computing} -- a curve-similarity
measure that, unlike DTW, preserves the monotonicity of the time axis. In addition, we varied the length of each temporal window around the baseline values detailed above (pre-grant: 8 and 9 years; during-grant: 4 and 5 years; post-grant: 5 and 7 years). Jointly, 17 additional clustering execution configurations were examined.

We assessed the statistical differences between the identified clusters using chi-square tests for gender distribution \cite{pearson1900criterion}
and one-way analysis of variance (ANOVA) with post-hoc Mann-Whitney U
tests with Bonferroni correction for continuous variables
\cite{fisher1992statistical,mann1947test,bonferroni1936teoria}. 
Comparing co-authorship activity across disciplines was performed using a Kruskal-Wallis test \cite{kruskal1952use} and Mann-Whitney U tests.
Statistical significance was set to $p = 0.05$.

\subsubsection{Machine learning prediction}
For the prediction of cluster membership based on researchers'
characteristics, we implemented three standard ML models: Logistic
Regression \cite{hosmer2013applied}, Random Forest
\cite{breiman2001random,shami2024implementing}, and XGBoost \cite{chen2016xgboost}. To
ensure strict temporal separation between predictors and outcomes, we
defined two separate prediction tasks: (1)~from the pre-grant to the
during-grant period (Pre~$\rightarrow$~During), and (2)~from the
during-grant to the post-grant period (During~$\rightarrow$~Post). All
bibliometric features (i.e., h-index, i10-index, total
publications, total citations, open access rate, and average number of collaborators) were recomputed using only publications falling within the relevant feature period. For instance, in the Pre~$\rightarrow$~During model, a researcher's h-index was calculated exclusively based on papers published before the grant award, rather than from the researcher's cumulative career record. This design ensures that no predictor encodes information from the period used to define the target cluster.  Table~\ref{tab:predictor_timing} specifies the temporal measurement window for each predictor across both transition models. A follow-up bivariate analysis using U-tests \cite{mann1947test} was performed to compare the average h-index, i10-index, total publications, and total citations between teams that sustained co-authorship after the grant and those that did
not. 

\begin{table}[H]
\centering
\caption{Predictor variables and their temporal measurement windows
for each ML transition model.}
\label{tab:predictor_timing}
\begin{tabular}{ll}
\toprule
\textbf{Feature} & \textbf{Definition} \\
\midrule
Academic age      & Years since first publication    \\
h-index           & Citation-based impact            \\
i10-index         & Papers with $\geq$10 citations   \\
Total papers      & Publication count                \\
Total citations   & Aggregate citation count         \\
Open access rate  & Proportion of OA publications    \\
Avg.\ collaborators & Mean co-authors per paper      \\
Gender composition & Proportion of male researchers \\
\midrule
\multicolumn{2}{l}{\textit{Note: Pre-grant window = 10 years before award year;}} \\
\multicolumn{2}{l}{\textit{During-grant window = award year through year +6.}} \\
\multicolumn{2}{l}{\textit{All features computed separately for Israeli and German}} \\
\multicolumn{2}{l}{\textit{researchers within each grant team.}} \\
\bottomrule
\end{tabular}
\end{table}

The model training and testing implemented an 80-20 percent
split such that 80\% is used for model training and the remaining 20\% for the model's evaluation \cite{SHMUEL2025131337}. Model performance was evaluated using
the K-fold stratified cross-validation method with $k=5$
\cite{diamantidis2000unsupervised}. Results are compared against a
majority-class (Zero-R) baseline that always predicts the most
frequent cluster, providing a minimal performance reference, using chi-square tests.

For each model, we conducted a feature importance analysis
\cite{zien2009feature,teddyfeature2024} by incorporating an additional
noise variable drawn from a standard normal distribution ($\mu=0,
\sigma=1$). Feature importance scores below the noise variable's score
were set to zero, and the remaining features underwent $L_1$
normalisation to enable comparison across algorithms.

\section{Results}
\label{sec:results}

First, we provide descriptive statistics of the collaboration
dynamics. Next, we present the clustering analysis outcomes. Finally, we detail the ML-based predictions of cluster membership.

\subsection{Descriptive Statistics}

\begin{figure}[H]
    \centering
    \includegraphics[width=1\textwidth]{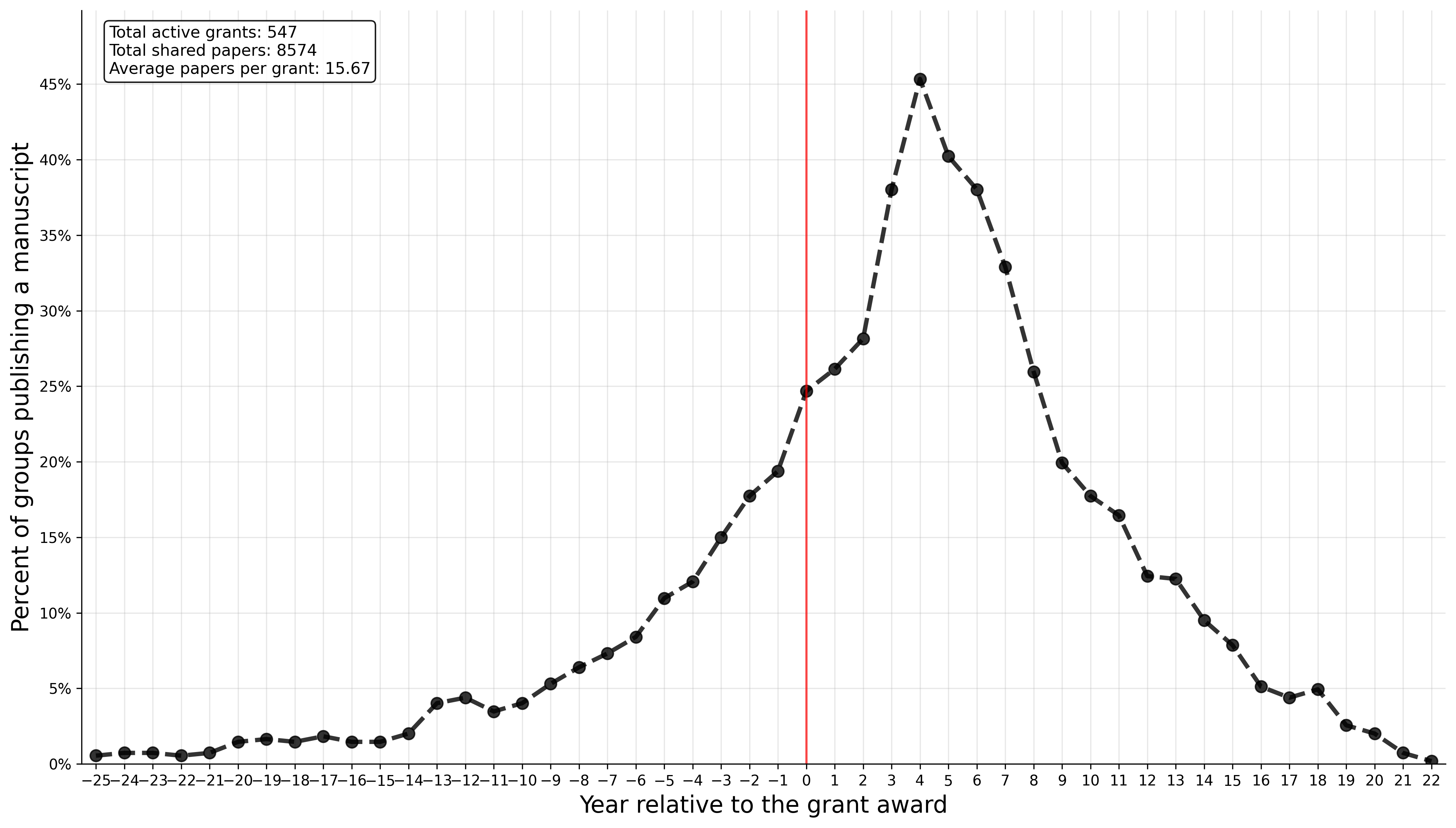}
    \caption{The temporal distribution of German--Israeli
    co-authorship rates within the GIF funding framework.
    The X-axis presents a 48-year range from 25 years pre-award year
    to 22 years post-award year, with year zero marking the award
    year, highlighted by a red dashed vertical line.}
    \label{fig:temporal_dist}
\end{figure}

\review{Out of the 642 grants analysed in this study, 100 (15.6\%) grants}
were not associated with any team co-authored publication, indicating
that these grants had no co-authorship at all before, during,
or after the grant period. Excluding these grants from this temporal
analysis, Figure~\ref{fig:temporal_dist} depicts the distribution of
the portion of German-Israeli co-authorship rates over time.
When focusing on the decade before and after the grant award year, the
data approximates a normal distribution over time with $p =
6\cdot10^{-4} < 0.05$, using a Wilk-Shapiro test
\cite{shapiro1965analysis}. The highest portion of co-authored
publications are presented 4 years after the funding commences, with
nearly half the teams (45.2\%) publishing a co-authored publication in
that year. Moreover, during the grant award year itself, 25\% of the
teams co-publish at least once. A wider pattern can also be observed
with the co-authorship ratio increasing, nearly
monotonically, in the range $t\in[-25, 4]$. Specifically, in the
decade prior to the grant award, the rate of co-authorship
is relatively low, with less than 5\% across the data. However, this
rate sharply increases towards $t=0$ and, in an almost symmetrical
manner, following $t=4$, it nearly monotonically decreases to the 5\%
range from $t=16$ onward.

\begin{figure}[H]
    \centering
    \includegraphics[width=1\textwidth]{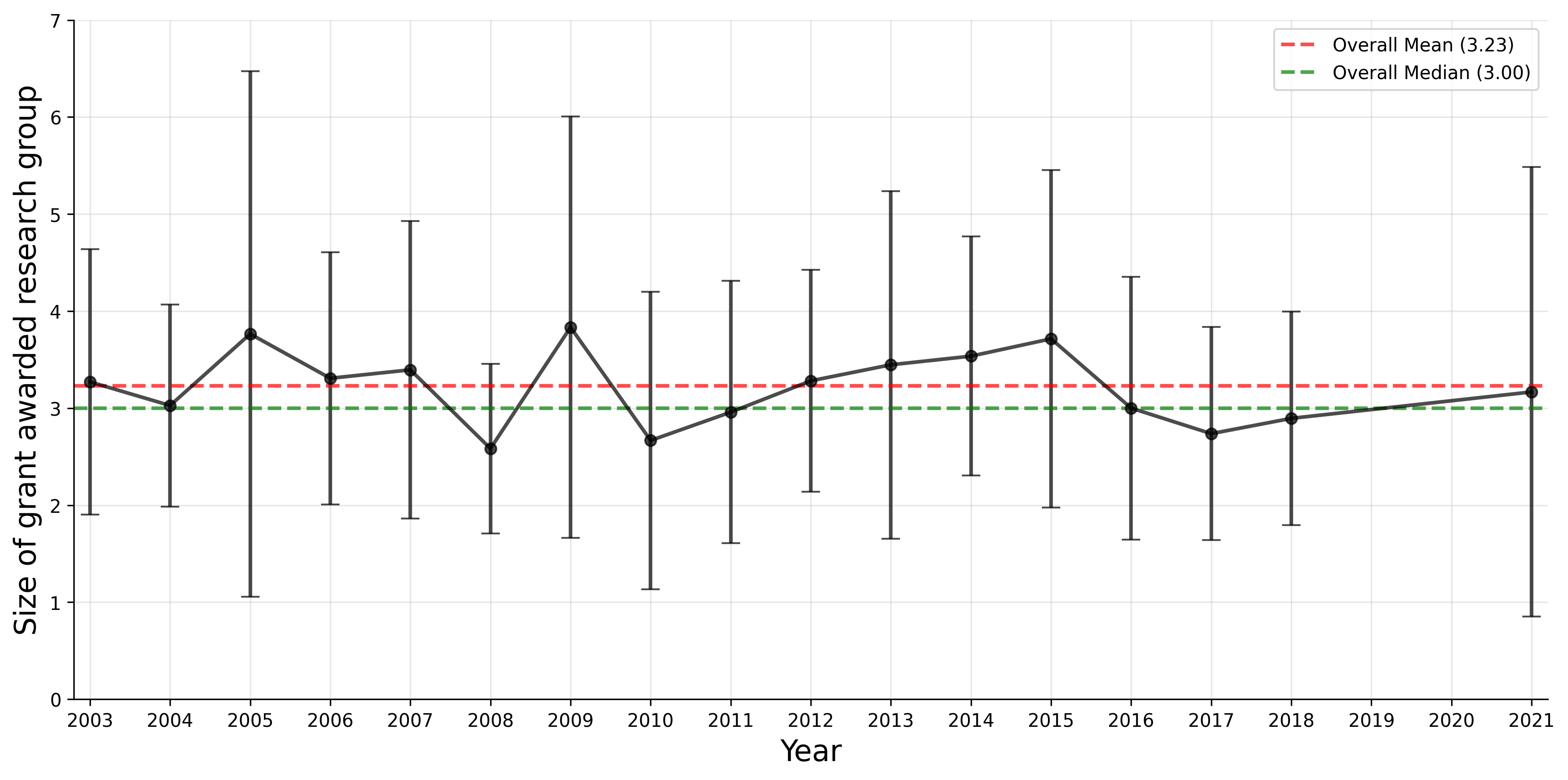}
    \caption{The temporal variation in the average number of
    researchers per awarded grant. Error bars represent one standard
    deviation. No data are available for 2019--2020 due to the
    absence of new awards during the COVID-19 pandemic.}
    \label{fig:researcher_per_grant}
\end{figure}

Figure~\ref{fig:researcher_per_grant} presents the temporal variation
in the average number of recipients per awarded grant. The error bars
indicate one standard deviation within the data. As we focus on
German-Israeli teams, it is safe to expect that all average values be
higher than 2, as indeed indicated in the figure. Overall, no clear
pattern can be observed with the number of recipients per grant
ranging between 2.5 and 4 team members (mean of 3.23, median of 3.00
researchers) throughout the years in question. It is important to note
that no data are available for the years 2019 and 2020 due to the
absence of new grant awards during the COVID-19 pandemic period
\cite{ciotti2020covid,lazebnik2021signature}. The error bars indicate
substantial variation within each year, suggesting considerable
diversity in team size across individual grants.

\begin{figure}[!ht]
    \centering
    \begin{subfigure}[t]{0.45\textwidth}
        \centering
        \includegraphics[width=\textwidth]{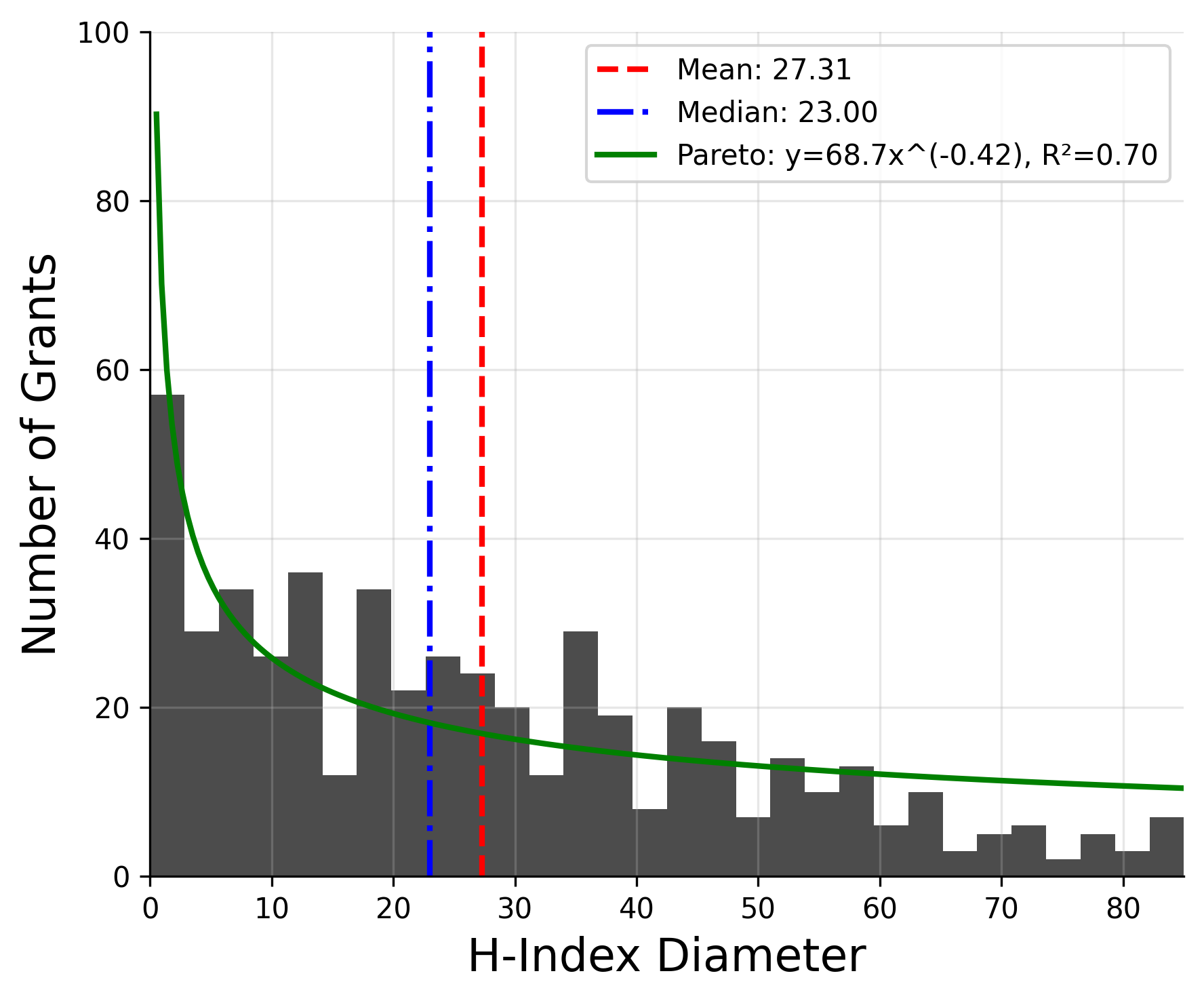}
        \caption{Distribution of h-index diameters.}
        \label{fig:subim1}
    \end{subfigure}
    \hfill
    \begin{subfigure}[t]{0.45\textwidth}
        \centering
        \includegraphics[width=\textwidth]{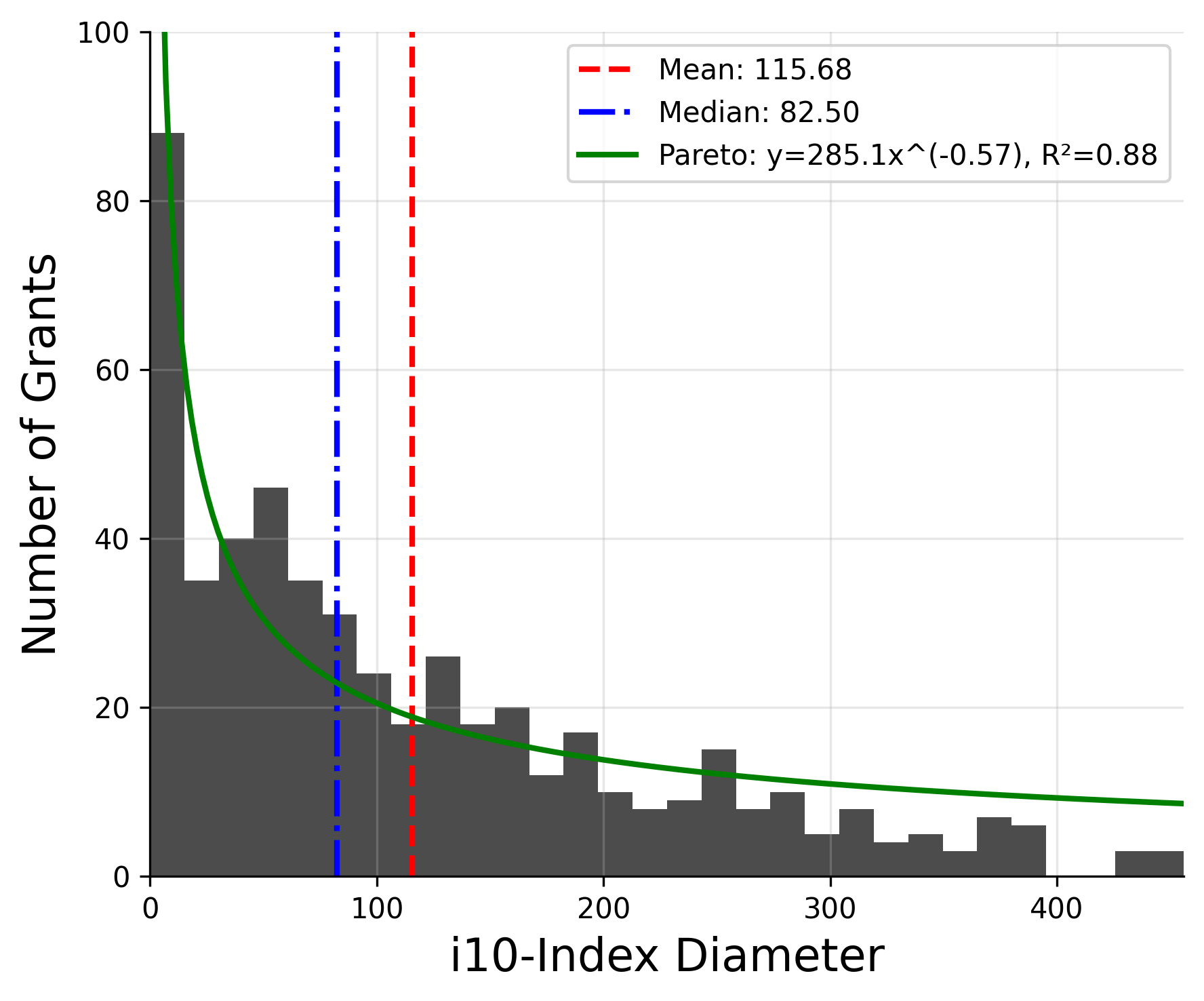}
        \caption{Distribution of i10-index diameters.}
        \label{fig:subim2}
    \end{subfigure}
    \\[1ex]
    \begin{subfigure}[t]{0.45\textwidth}
        \centering
        \includegraphics[width=\textwidth]{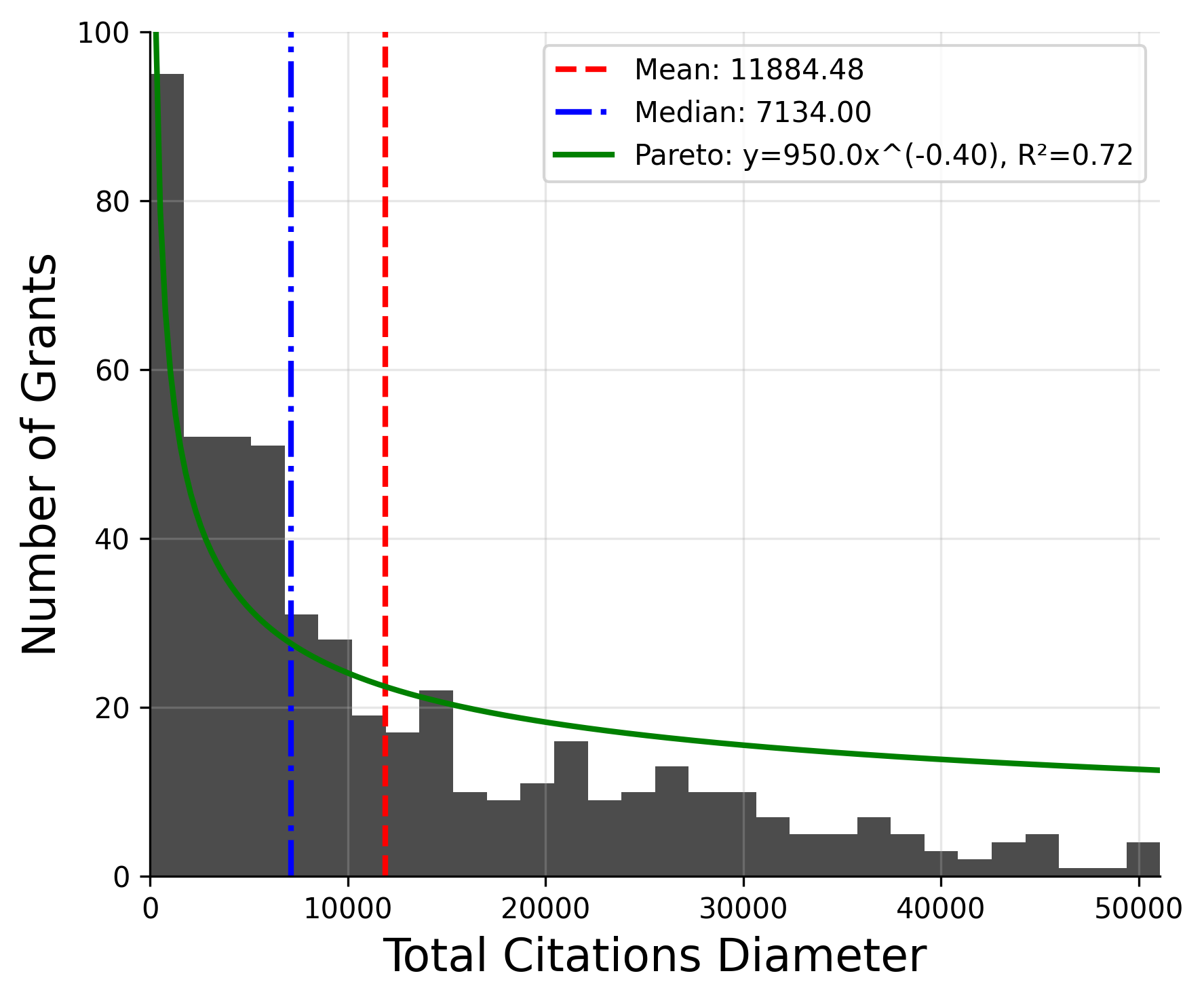}
        \caption{Distribution of total citation diameters.}
        \label{fig:subim3}
    \end{subfigure}
    \hfill
    \begin{subfigure}[t]{0.45\textwidth}
        \centering
        \includegraphics[width=\textwidth]{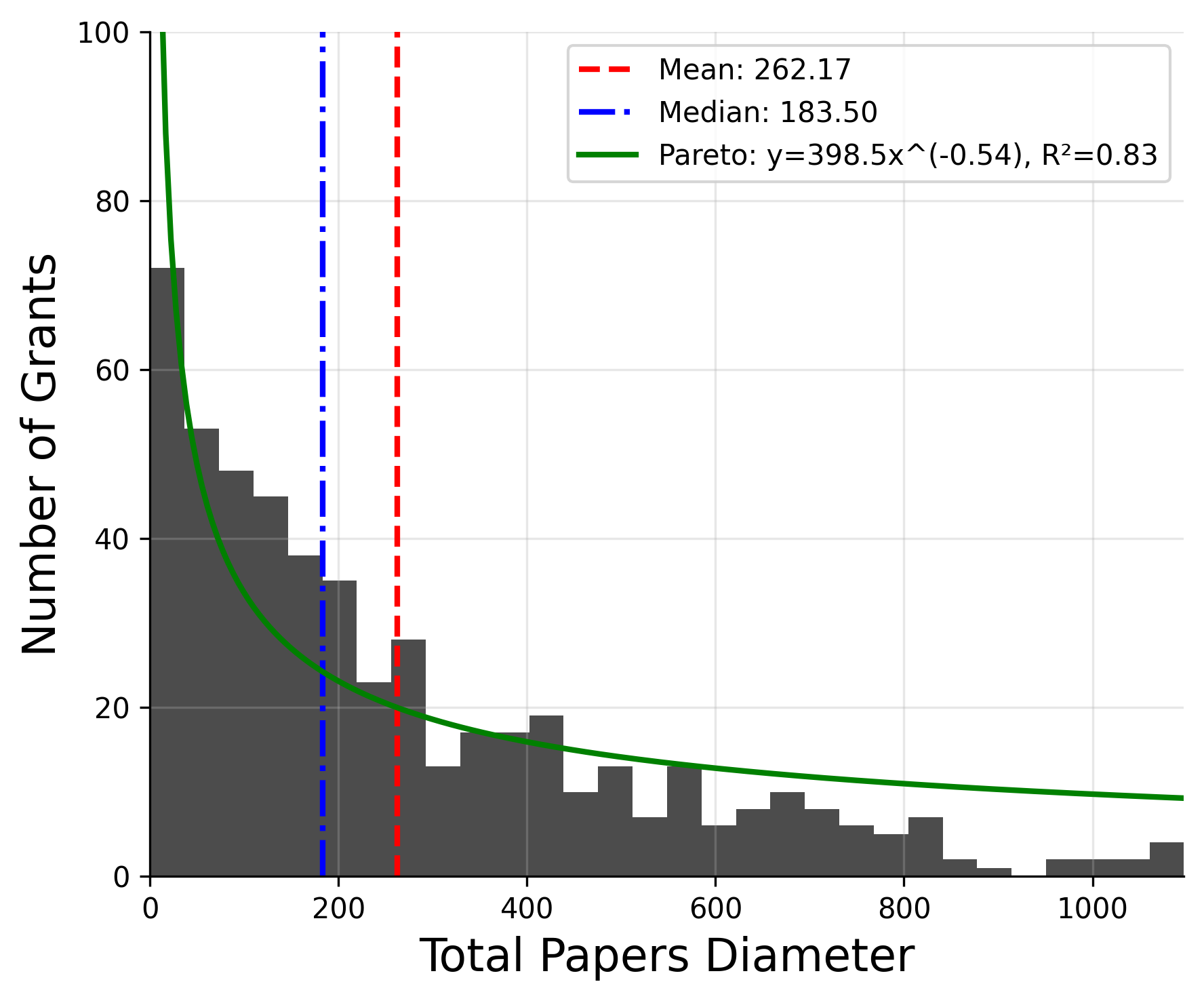}
        \caption{Distribution of total publication diameters.}
        \label{fig:subim4}
    \end{subfigure}
    \caption{Distributions of academic metric diameters across
    research teams supported by GIF grants. All four measures exhibit
    right-skewed distributions with Pareto-like tails.}
    \label{fig:mainfig}
\end{figure}

Figure~\ref{fig:mainfig} illustrates the distribution of key
bibliometric diameters across research teams within the GIF funding
scheme. A bibliometric diameter is defined as the largest difference
in a given metric, $M$, between any two members $s_i, s_j$ of a team
$s_i, s_j \in S$. Formally, the diameter of metric $M$, denoted by
$D_M$ over group $S$, is defined as $D_M(S) := \max\big(\{\forall
s_i, s_j \in S: |M(s_i)-M(s_j)|\}\big)$. Specifically,
panel~\ref{fig:subim1} presents the distribution of h-index diameters,
panel~\ref{fig:subim2} focuses on i10-index diameters,
panel~\ref{fig:subim3} reports the corresponding results for total
citation counts, and panel~\ref{fig:subim4} for total publication
counts. Across all four metrics, the distributions are markedly
right-skewed, with strong Pareto distribution fitting ($R^2$ ranging
between 0.70 and 0.88). These underscore the prevalence of relatively
homogeneous teams alongside a smaller subset with substantial
intra-team disparities. The consistent pattern of means exceeding
medians (h-index: 27.31 vs.\ 23.00; i10-index: 115.68 vs.\ 82.50;
citations: 11,884.48 vs.\ 7,134.00; publications: 262.17 vs.\ 183.50)
further highlights the presence of a long tail of highly heterogeneous
teams. Taken together, panels~\ref{fig:subim1}--\ref{fig:subim4}
demonstrate that while many grants are awarded to teams of comparable
researchers' standing, a non-negligible part thereof is awarded to
teams with sharply contrasting academic standings. Supplementary analysis of open-access publishing patterns 
among GIF-funded researchers is provided in 
Appendix~\ref{app:Open-Access}.

\subsection{Trend and Survival Analysis}

Linear regression on the pre-grant
decade ($t\in[-10,-1]$) revealed a significant positive trend
($\beta = 0.017$, $R^2 = 0.97$, $p < 0.001$), confirming that
co-authorship rates rise consistently in the years leading up to the
grant. On the post-grant decade ($t\in[6,16]$), a significant negative
trend of nearly twice the magnitude was found ($\beta = -0.030$,
$R^2 = 0.93$, $p < 0.001$), indicating a sharper decline after the
grant concludes than the rise that preceded it. The two slopes differ significantly ($Z = 15.74$, $p < 0.001$),
establishing an asymmetry between co-authorship formation
and dissolution around the funding period. Among the 451 teams active
during the grant period, the survival analysis revealed a median post-grant
co-authorship survival time of just two years, with 450 out of 451
teams eventually discontinued co-authorship within a decade.

\subsection{Clustering Analysis}

Our clustering analysis revealed three main clusters (see Appendix for
the $K$ value search graph). By manual investigation of the
algorithmically derived cluster centroids, we chose to name these
clusters as \say{No co-publications cluster}, \say{Several
co-publications cluster}, and \say{High-volume publications cluster},
as the number of co-publications arises as a straightforward indicator
to distinguish between them. Namely, the first cluster is associated
with no co-publications at all, while the second and third clusters
are associated with less than or equal to 3 co-publications per year
on average, and more than 3 co-publications per year on average,
respectively. Importantly, these labels and the associated
thresholds are descriptive characterisations of the data-driven
groupings; no predefined cutoffs were imposed on the clustering
procedure.

Table~\ref{tab:academic_metrics_full} presents a comparison between
the three clusters in the researcher-level bibliometrics across
clusters in the pre-grant, during-grant, and post-grant periods. One
can notice that teams with no pre-grant co-publications are not
necessarily composed of early-career researchers. In fact, these teams
display substantial productivity and impact, with an average h-index
of 46.2 and more than 280 publications per researcher, roughly
comparable with teams that already had several pre-grant
co-publications. By contrast, high-volume pre-grant co-publications
consist of more experienced and productive researchers, reflected by
their higher average h-index (54.2), i10-index (171.4), and total
output (398.5 publications on average). During the grant period,
moderate and high-volume co-publication clusters are both
characterised by higher average bibliometric scores than the no
co-publication teams. High-volume teams, for example, average more
than 345 publications and over 15,000 citations per researcher,
compared to 225 publications and 10,027 citations for inactive teams.
The average number of researchers per grant also increases with
collaboration intensity, rising from about 3 for inactive or
moderately productive teams to above 4 in high-output collaborations.
In the post-grant period, academic age naturally increases across all
clusters, but collaboration intensity continues to differentiate teams.
High-volume post-grant groups again display the strongest metrics
(h-index 52.7, i10-index 162.9, nearly 16,100 citations), and they
also sustain the largest team sizes (4.68 researchers per grant). By
contrast, inactive post-grant teams maintain lower bibliometric
indicators despite similar academic age. Gender distribution remains
stable across all clusters and periods, with male ratios consistently
around 0.80--0.85.

Chi-square tests revealed no significant differences in gender
distribution across clusters in any of the three time periods
(pre-grant: $\chi^2 = 1.29$, $p = 0.525$; during grant:
$\chi^2 = 2.97$, $p = 0.227$; post-grant: $\chi^2 = 2.70$,
$p = 0.259$). Moreover, an ANOVA test presented statistically
significant differences between clusters across all bibliometric
measures in each time period. Post-hoc U tests show all pairs are
also statistically significantly different.

\begin{figure}[ht]
    \centering
    \includegraphics[width=1\textwidth]{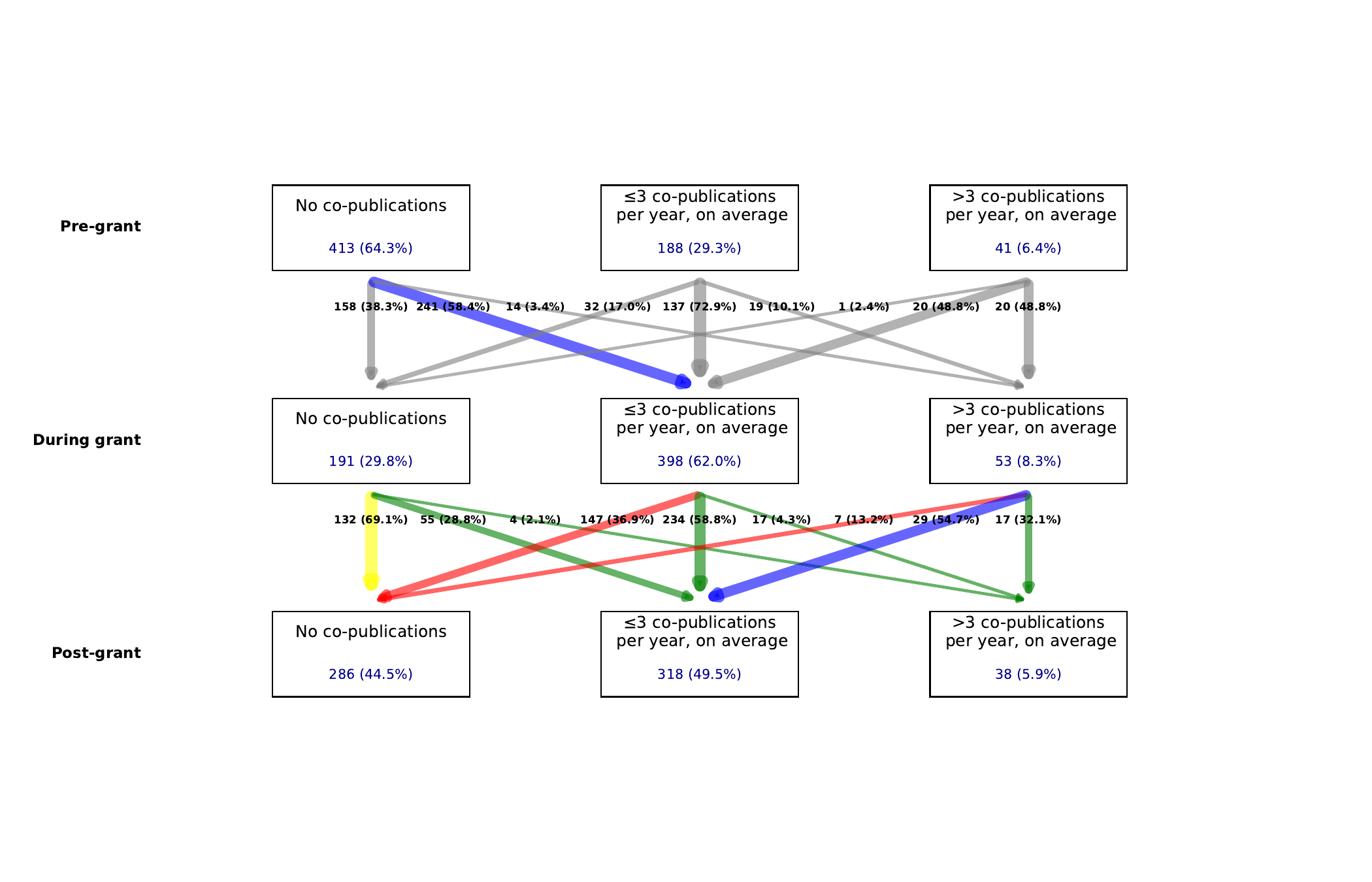}
    \caption{Temporal cluster analysis of the co-authorship
    dynamics across pre-grant, during-grant, and post-grant periods.
    The colours highlight prototypical trajectories: blue --
    funding-oriented dynamics; green -- successful sustained
    co-authorship; red -- collapse after funding; yellow --
    persistently inactive teams; gray -- other transitions.}
    \label{fig:transitions}
\end{figure}

Figure~\ref{fig:transitions} illustrates the dynamics of collaboration
intensity between three phases: pre-grant, during-grant, and
post-grant. Each box represents the proportion of teams associated
with each cluster. Arrows capture how teams transition across clusters
between periods, with colours highlighting key types of trajectories:
\textbf{blue} arrows denote funding-oriented activation paths where
previously inactive or moderately active teams increase collaboration
when funding becomes available, and decrease it when funding stops;
\textbf{green} arrows represent seemingly desired transitions, where
co-authorship is sustained or even strengthened beyond the
funding period; \textbf{red} arrows mark collapse paths, where
co-authorship ceases entirely once funding ends; and
\textbf{yellow} arrows denote persistently inactive groups that fail
to co-publish during and even after the funding. Gray arrows capture
all other transitions.

Starting with the pre- to during-grant transition, we see a strong
funding-oriented transition. Notably, 241 out of 413 grants (58.4\%)
with no prior co-publication became collaborative ($\leq 3$
co-publications per year, on average) during the grant period (blue).
An additional 14 grants (3.4\%) leaped directly from inactivity to
high output ($>3$ publications per year, on average). In contrast, 32
out of 188 moderately active pre-grant teams (17.0\%) produced no
outputs during the grant, indicating occasional reversals. High-output
teams before the grant (41 in total) largely maintained their
co-authorship, with 20 (48.8\%) remaining highly productive
and another 20 (48.8\%) reducing to moderate levels. The transition
from the grant period to the post-grant period reveals the
sustainability of these co-authorship ties. Here, 234 out of
398 (58.8\%) moderately collaborative teams remained so, and 17
(4.3\%) even advanced to high productivity (green). Among the most
productive teams during the grant, 17 out of 53 (32.1\%) remained
highly productive post-grant, while 29 (54.7\%) lowered to moderate
levels (blue). However, a substantial fraction of
co-authorship ties collapsed once funding ended. Specifically,
147 out of 398 (36.9\%) moderately collaborative groups and 7 out of
53 (13.2\%) highly collaborative teams produced no co-publications
after the grant (red). At the same time, 132 out of 191 groups
(69.1\%) that were already inactive during the grant remained inactive
afterward (yellow), representing missed opportunities where funded
projects failed to spark co-authorship. Overall, more than
half (58.4\%) of previously inactive teams became active during the
funding period. Moreover, sustainability is limited: almost
one-quarter of all co-authorship ties (154 grants, 24.0\%)
collapsed after the grant ended. Furthermore, long-term success is
uneven: while about one-third of high-performing teams sustained their
co-authorship, nearly 70\% of inactive teams never
co-publish.

To further examine whether post-grant co-authorship
persistence varies by prior collaboration rates, we compared the
pre-grant cluster membership of teams that sustained co-authorship
after the grant against those that did not. Among teams that were
active during the grant period ($n = 452$), 65.9\% maintained at
least some co-authorship activity post-grant, while 34.1\% collapsed
entirely. Persistent teams had a
significantly higher pre-grant collaboration rates than
non-persistent teams ($U = 26990.5$, $p < 0.001$). Specifically, among teams with no prior
co-authorship (pre-grant), 59.6\% remained active
post-grant, compared to 70.5\% of moderately active pre-grant teams
 and 87.5\% of highly active pre-grant teams.
This suggests that prior collaborative ties are a significant
predictor of post-grant co-authorship persistence.

To examine whether co-authorship dynamics vary across
disciplinary contexts, we grouped the grants into three broad categories: STEM (Physics, Chemistry, Mathematics, Computer Sciences, Electrical Engineering,
Mechanical Engineering, and Engineering Sciences; $n = 210$), Life
Sciences (Biology, Medicine, Biochemistry, Agricultural Sciences,
Practical Medicine, Geology, and Environmental Sciences; $n = 159$),
and Humanities and Social Sciences (Psychology, Economics, Law,
History, Social Sciences, Language Sciences, Geography, Old and
Oriental Culture, and History of Sciences; $n = 73$).  Significant
differences across disciplines in co-authorship activity were observed during the
grant period ($H = 13.13$, $p = 0.001$), with STEM teams showing the
highest active co-authorship rate (89.0\%), followed by Life Sciences
(82.4\%) and Humanities and Social Sciences (68.5\%). A chi-square
test similarly confirmed significant differences in cluster
distribution during the grant period ($\chi^2 = 17.34$, $p = 0.002$). However, post-grant persistence rates
did not differ significantly across disciplines ($\chi^2 = 4.19$,
$p = 0.382$), with rates of 70.6\%, 71.0\%, and 86.0\% for STEM,
Life Sciences, and Humanities and Social Sciences respectively.

The robustness analysis, conducted across all 17 configurations, 
consistently recovered three clusters whose composition and transition
patterns closely mirror those reported in our main analysis. Specifically, 
in all configurations the optimal partition comprised the same three 
qualitative groups -- inactive, moderately active, and highly active teams 
-- with cluster sizes remaining broadly stable across configurations. Full results 
for all configurations are reported in Appendix~\ref{app:robustness}.

\subsection{Machine Learning Prediction for Cluster Association}

Table~\ref{tab:model_performance} presents the predictive
performance of each model for both examined transitions (Pre~$\rightarrow$~During and During~$\rightarrow$~Post), alongside a
majority-class (Zero-R) baseline. For the Pre~$\rightarrow$~During
transition, XGBoost achieves the most favorable performance (accuracy = 0.67), followed by Random Forest (accuracy = 0.51)
and Logistic Regression (accuracy = 0.34). For the During~$\rightarrow$~Post transition, XGBoost brings about the highest performance (accuracy = 0.60), followed by Random Forest (accuracy =
0.54) and Logistic Regression (accuracy = 0.36). Crucially, across both
transitions, the best-performing ML model (XGBoost) does not outperform the majority-class baseline in accuracy. In fact, the baseline model achieves significantly higher accuracy than the Logistic Regression (both transitions) and the Random Forest (Pre~$\rightarrow$~During). 

A follow-up bivariate analysis of the average h-index, i10-index, total publications, and total citations between teams that sustained co-authorship after the grant and those that did
not revealed no statistically significant
differences. 

A feature analysis of the ML models are provided in Appendix ~\ref{app:feature_analysis}.

\section{Discussion}
\label{sec:discussion}

Our results combine to suggest that while bi-national funding,
exemplified by the GIF, is associated with increased
collaborative activity during the grant period, this
association rarely extends to new, lasting partnerships between the
researchers involved. As Figure~\ref{fig:temporal_dist} demonstrates,
co-authorship activity rises sharply slightly before and mainly
during the grant period but declines rapidly afterward,
returning to pre-grant levels within a decade. Similarly, the cluster
flow analysis and Table~\ref{tab:academic_metrics_full} show that most
\say{new} co-authorship ties (i.e., those with no prior
co-publications) either dissolve entirely or regress to very low
levels of co-authorship activity once funding ends. Thus, rather than being
indicative of enduring collaborative partnerships, GIF-supported
collaborations often appear as temporary alliances bounded by the
grant cycle. In other words, the GIF appears to be associated
with reinforcing existing ties and short-lived bursts of productivity, but
is rarely associated with self-sustaining partnerships.

These findings align with prior work on the fragility of
grant-induced co-authorship ties across funding scales.
At the multinational level, studies of the EU Framework Programmes
have documented similar patterns of post-funding attrition
\cite{Morillo2019Collaboration}.
Comparable dynamics have been observed under national funding
schemes in diverse contexts, including competitive research grants
\cite{jacob2011impact,c2,saygitov2014impact},
institutional incentive programmes \cite{schroen2012research},
and broader analyses of funding-induced collaboration networks
\cite{c1,c3}.
Additionally, network theory suggests that durable ties require repeated, multi-contextual interactions that build trust, shared infrastructure, and mutual
dependency \cite{c5}. However, by design, bi-national partnerships
are dyadic, small-scale, and project-based. Without larger consortia
or structural incentives to embed collaborations within institutional
frameworks, partnerships lack resilience once
the financial \say{glue} is removed \cite{c6}. Moreover, the skewed
distribution of differences in recipients' bibliometrics, as depicted
in Figure~\ref{fig:mainfig}, indicates that many teams pair
researchers with very different levels of experience and standing.
While such asymmetry may yield short-term complementarities, it is
often assumed to undermine the reciprocity needed for long-term
continuity \cite{alexi2024scientometrics}.

\review{Across both examined transition phases (pre-grant to during-grant and during-grant to post-grant), standard bibliometric indicators demonstrated negligible discriminative power: even the highest-performing model failed to achieve predictive accuracy meaningfully beyond a naive majority-class baseline. Therefore, this result is primarily diagnostic and negative, indicating that standard bibliometric profiles of team members are poor predictors of cluster membership. This is not evidence about the mechanisms underlying co-authorship persistence. One hypothesis consistent with this negative result, which our analysis cannot itself confirm, is that co-authorship trajectories are shaped by latent contextual and relational factors that are not captured by standard bibliometrics. This interpretation is consistent with prior work suggesting that researcher mobility and structural enablers are associated with more durable international networks \cite{scellato2015migrant,franzoni2017mobility} than individual researcher characteristics or isolated project-based funding \cite{bozeman2004scientists,c2}.}

Taken jointly, the GIF appears to be associated with
short-term collaborative activity, yet the bibliometric data do not support the persistence of these activities. Recognizing this limitation opens the door for improving and even redesigning bi-national funding schemes in ways that move beyond temporary alliances and instead nurture resilient, long-lasting
networks. With appropriate adjustments, such programmes can better
align scientific collaboration with broader policy goals.

It is important to note that this study is not without limitations. First, we adopted the common
bibliometric approach, where co-authorship is considered as a proxy for
collaboration \cite{subramanyan1983bibliometric}. This approach
overlooks other important forms of partnership, such as joint grant
applications, student exchanges, or shared data \cite{kahn2018co}.
In addition, our reliance on OpenAlex as the sole bibliometric
data source introduces potential coverage biases, as certain
publication types, languages, or regional journals may be
underrepresented in the database \cite{priem2022openalex}.
Therefore, future research could combine additional bibliometric data with surveys,
interviews, and mobility records to capture a more nuanced perspective.
Second, our design cannot fully disentangle whether GIF funding creates
new collaborations or merely reinforces existing ones, suggesting that
causal evaluation methods, such as matched controls or regression
discontinuity designs, would strengthen future analyses
\cite{pearce2016analysis,imbens2008regression}. Furthermore,
while our disciplinary analysis reveals that STEM teams are
significantly more likely to co-author publications during the grant period than
Humanities and Social Sciences teams, post-grant survival rates do
not differ significantly across disciplines. This suggests that disciplinary
differences primarily shape the likelihood of initiating co-authorship
rather than its durability once established. Nevertheless,
field-normalised metrics and non-traditional outputs such as software,
patents, or policy briefs, remain outside the scope of the current
study \cite{alperin202213,bantilan2004dealing}, and future work should
incorporate these dimensions for a more complete picture.

\section{Conclusions}
\label{sec:conclusions}

In this study, we explored the relationship between bi-national academic funding and long-term collaboration dynamics using the German-Israeli Foundation (GIF) as a case study. By tracking co-authorship trajectories across 642 grants and integrating temporally-aware clustering with machine learning, we demonstrated that \review{while bi-national funding is reliably associated with elevated collaborative activity during the grant period, this activity rarely persists as enduring co-authorship}. Co-authorship typically spikes around the funding window but regresses to pre-grant levels shortly after.\review{Furthermore, our predictive modeling revealed that standard bibliometric indicators fail to forecast post-grant collaboration persistence; we hypothesise — though our analysis cannot establish this — that long-term scientific ties are shaped more by contextual and relational factors than by bibliometric ones.} These findings contribute to the broader research policy literature by challenging the assumption that dyadic, project-based funding inherently forges sustainable international networks.

Consequently, for funding agencies and research institutions, these insights highlight the potential limitations of relying solely on isolated funding cycles to overcome project-based attrition. If the goal is to build resilient networks, alternative structural arrangements must be conceptualised and empirically tested. Potential mechanisms to investigate could include sequential funding opportunities, institutional anchoring (such as joint research centers or shared infrastructure), and multi-nodal consortia that extend beyond dyadic pairings. Furthermore, given our finding that bibliometrics are an insufficient predictor of durable partnerships, alternative evaluation criteria warrant exploration. Shifting evaluation frameworks to reward signals of embedded cooperation -- such as sustained trainee exchange or joint follow-up applications -- provides a promising theoretical alternative that should be rigorously evaluated in future research.

Future investigations should also expand upon these descriptive foundations to build a more comprehensive model of network resilience. First, comparative studies across multiple funding programs -- contrasting bi-national schemes with larger multi-national or strictly national initiatives -- would help isolate the structural mechanisms that best support long-term collaboration. Second, the methodological reliance on co-authorship should be broadened to include additional collaborative indicators, such as joint grant applications, shared datasets, and patent co-inventions. Finally, a deeper exploration of disciplinary differences is warranted; while our findings suggest that STEM fields initiate co-authorship more readily during the grant window, understanding the specific mechanisms that allow different disciplines to sustain these ties over time could inform more tailored, domain-specific funding strategies.

\section*{Declarations}

\subsection*{Funding}
No funding was received to assist with the preparation of this
manuscript.

\subsection*{Conflicts of interest/Competing interests}
The authors have no competing interests to declare that are relevant
to the content of this article.

\subsection*{Code and Data availability}
The code and data that have been used in this study are available from the corresponding author.

\subsection*{Author Contribution}
Amit Bengiat: Conceptualization, Methodology, Software, Formal
analysis, Data Curation, Writing -- Original Draft, Visualization. \\
Teddy Lazebnik: Software, Supervision, Formal analysis,
Investigation, Validation, Resources, Writing -- Review \& Editing,
Visualization. \\
Philipp Mayr: Validation, Writing -- Review \& Editing. \\
Ariel Rosenfeld: Conceptualization, Supervision, Methodology,
Validation, Writing -- Original Draft, Writing -- Review \& Editing.

\bibliographystyle{plain}
\bibliography{references}

\section*{Appendix}

\subsection*{Cluster Optimal Size}

\begin{figure}[H]
    \centering
    \includegraphics[width=0.99\linewidth]{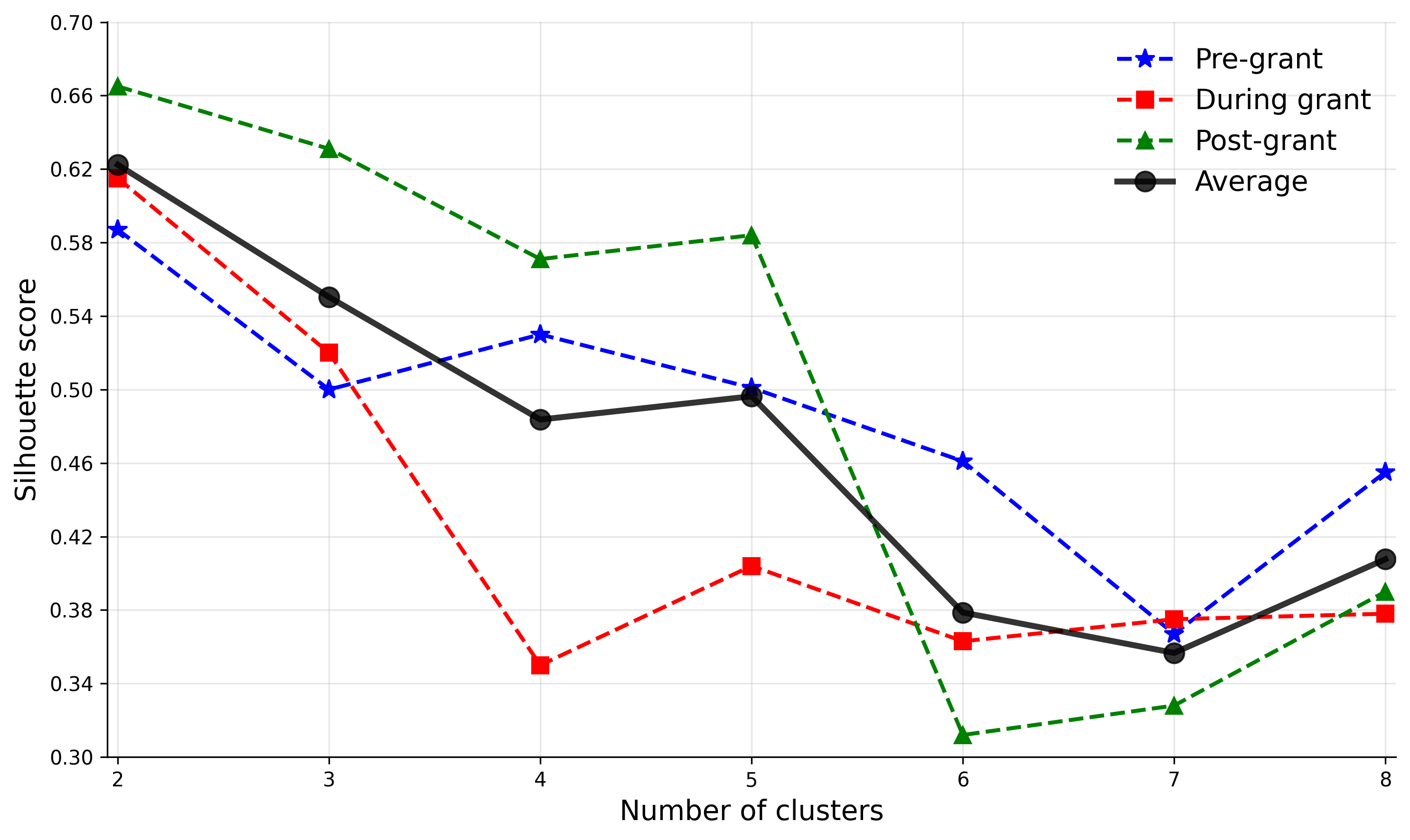}
    \caption{Silhouette score over different number of clusters ($K$).}
    \label{fig:pick_k_clusters}
\end{figure}

Figure~\ref{fig:pick_k_clusters} presents the silhouette scores
obtained for different numbers of clusters ($K$) across the three
temporal phases. Despite some variation between the pre-grant,
during-grant, and post-grant periods, the average score across all
phases shows a clear global maximum at $K=3$. This indicates
that three clusters provide the most consistent and robust partitioning
of the data when considering all periods jointly. Although higher
values of $K$ occasionally yield locally improved scores in specific
periods (e.g., post-grant at $K=5$), these solutions are not stable
across the full temporal scope and result in a marked drop in the
average silhouette score. By contrast, the three-cluster solution
balances interpretability with statistical validity, offering a
coherent structure that captures the main collaboration trajectories
without overfitting.

\begin{table*}[!ht]
\centering
\caption{Average research-level bibliometrics for different clusters
during the pre-grant, during-grant, and post-grant periods.}
\label{tab:academic_metrics_full}
\begin{adjustbox}{width=\textwidth,center}
\begin{tabular}{l c S[table-format=3.0] S[table-format=4.0]
  S[table-format=2.1,separate-uncertainty]
  S[table-format=2.1,separate-uncertainty]
  S[table-format=3.1,separate-uncertainty]
  S[table-format=5.1,separate-uncertainty]
  S[table-format=4.1,separate-uncertainty]
  S[table-format=1.2]}
\hline\hline
\multicolumn{1}{c}{\bfseries Time Frame} &
\multicolumn{1}{c}{\bfseries Cluster} &
\multicolumn{1}{c}{\bfseries Total grants} &
\multicolumn{1}{c}{\bfseries Total researchers} &
\multicolumn{1}{c}{\bfseries Academic age} &
\multicolumn{1}{c}{\bfseries h-index} &
\multicolumn{1}{c}{\bfseries i10-index} &
\multicolumn{1}{c}{\bfseries Total citations} &
\multicolumn{1}{c}{\bfseries Total papers} &
\multicolumn{1}{c}{\bfseries Avg researchers per grant} \\
\hline\hline
\multirow{3}{*}{\shortstack{Pre-grant\\(10 years)}}
& No co-publications
  & 413 & 900 & 17.2 $\pm$ 21.5 & 46.2 $\pm$ 28.5 & 130.5 $\pm$ 138.0
  & 12484.2 $\pm$ 16830.1 & 285.5 $\pm$ 308.1 & 2.88 \\
& $\leq$ 3 co-pub/year
  & 188 & 588 & 22.5 $\pm$ 27.4 & 44.0 $\pm$ 27.3 & 128.2 $\pm$ 135.2
  & 11436.7 $\pm$ 15095.5 & 283.1 $\pm$ 282.8 & 3.13 \\
& $>$ 3 co-pub/year
  & 41  & 190 & 19.5 $\pm$ 22.4 & 54.2 $\pm$ 33.3 & 171.4 $\pm$ 170.7
  & 17001.8 $\pm$ 23246.9 & 398.5 $\pm$ 400.9 & 4.63 \\
\midrule
\multirow{3}{*}{\shortstack{During grant\\(6 years)}}
& No co-publications
  & 191 & 268 & 29.4 $\pm$ 26.1 & 38.8 $\pm$ 27.2 & 101.5 $\pm$ 114.4
  & 10027.2 $\pm$ 14715.6 & 225.8 $\pm$ 232.6 & 2.95 \\
& $\leq$ 3 co-pub/year
  & 398 & 1138 & 29.1 $\pm$ 23.7 & 45.7 $\pm$ 27.2 & 130.4 $\pm$ 133.1
  & 11931.3 $\pm$ 15473.3 & 285.2 $\pm$ 297.0 & 2.86 \\
& $>$ 3 co-pub/year
  & 53  & 215 & 28.4 $\pm$ 22.0 & 50.4 $\pm$ 30.5 & 153.7 $\pm$ 169.1
  & 15008.4 $\pm$ 22020.7 & 345.7 $\pm$ 379.0 & 4.06 \\
\midrule
\multirow{3}{*}{\shortstack{Post-grant\\(5 years)}}
& No co-publications
  & 286 & 562 & 36.0 $\pm$ 27.2 & 44.7 $\pm$ 25.6 & 122.8 $\pm$ 121.8
  & 11299.7 $\pm$ 14002.6 & 261.6 $\pm$ 246.6 & 3.02 \\
& $\leq$ 3 co-pub/year
  & 318 & 965 & 33.3 $\pm$ 22.6 & 45.2 $\pm$ 28.8 & 130.4 $\pm$ 138.7
  & 12197.4 $\pm$ 17118.0 & 290.7 $\pm$ 314.5 & 3.03 \\
& $>$ 3 co-pub/year
  & 38  & 178 & 35.2 $\pm$ 21.9 & 52.7 $\pm$ 30.1 & 162.9 $\pm$ 166.0
  & 16087.2 $\pm$ 22873.4 & 368.2 $\pm$ 377.0 & 4.68 \\
\hline\hline
\end{tabular}
\end{adjustbox}
\end{table*}

\begin{table*}[!ht]
\centering
\caption{Predictive performance of ML models and the
majority-class baseline for two cluster transitions. All bibliometric
features were computed exclusively from publications within the
relevant temporal window.}
\label{tab:model_performance}
\begin{adjustbox}{width=\textwidth,center}
\begin{tabular}{llccccc}
\hline\hline
\textbf{Transition} & \textbf{Algorithm} & \textbf{Accuracy}
& \textbf{F1 Score} & \textbf{Recall} & \textbf{Precision}
& \textbf{AUC} \\
\hline\hline
\multirow{4}{*}{Pre $\rightarrow$ During}
& Majority class (Zero-R)
  & 0.73 & 0.61 & -- & --
  & 0.50 \\
& Logistic Regression
  & 0.34 & 0.39 & 0.34 & 0.51
  & 0.43 \\
& Random Forest
  & 0.51 & 0.51 & 0.51 & 0.52
  & 0.47 \\
& XGBoost
  & 0.67 & 0.62 & 0.67 & 0.61
  & 0.58 \\
\midrule
\multirow{4}{*}{During $\rightarrow$ Post}
& Majority class (Zero-R)
  & 0.57 & 0.42 & -- & --
  & 0.50 \\
& Logistic Regression
  & 0.36 & 0.40 & 0.36 & 0.49
  & 0.47 \\
& Random Forest
  & 0.54 & 0.54 & 0.54 & 0.53
  & 0.56 \\
& XGBoost
  & 0.60 & 0.60 & 0.60 & 0.61
  & 0.66 \\
\hline\hline
\end{tabular}
\end{adjustbox}
\end{table*}

\subsection*{Clustering Robustness}
\label{app:robustness}

To verify that the three-cluster solution and the resulting
co-authorship trajectories are not artefacts of a particular
methodological choice, we conducted a systematic robustness analysis
across 17 configurations. These configurations varied along two
dimensions: (i)~the distance metric -- Dynamic Time Warping (DTW)
\cite{sakoe1978dynamic} versus the discrete Fr\'{e}chet distance
\cite{eiter1994computing,alt1995computing} -- and (ii)~the length of
each temporal window (pre-grant: 8, 9, or 10 years; during-grant: 4,
5, or 6 years; post-grant: 5 or 7 years). In every configuration the
same pipeline was applied: grants with zero co-publications in the
relevant windows were separated into a dedicated cluster, the remaining
grants were embedded via MDS on the pairwise distance matrix, and
K-means with silhouette-score optimisation was used to select the
number of active-grant clusters.

Across all 17 configurations, three qualitative findings remain
stable. First, the optimal number of clusters (excluding the
zero-publication group) is consistently two, yielding a total of
three clusters that map onto the same substantive categories reported
in the main text. Second, the dominant transition pattern is
preserved: a majority of previously inactive teams become moderately
active during the grant, yet a substantial share reverts to inactivity
once funding ends. Third, the proportion of teams that sustain or
strengthen co-authorship post-grant remains a minority across all
configurations. Taken together, these results confirm that the
trajectory typology and the associated policy conclusions are robust
to reasonable perturbations of the distance metric and temporal window
length.

\begin{figure}[H]
    \centering
    \begin{subfigure}[t]{0.48\textwidth}
        \centering
        \includegraphics[width=\textwidth]{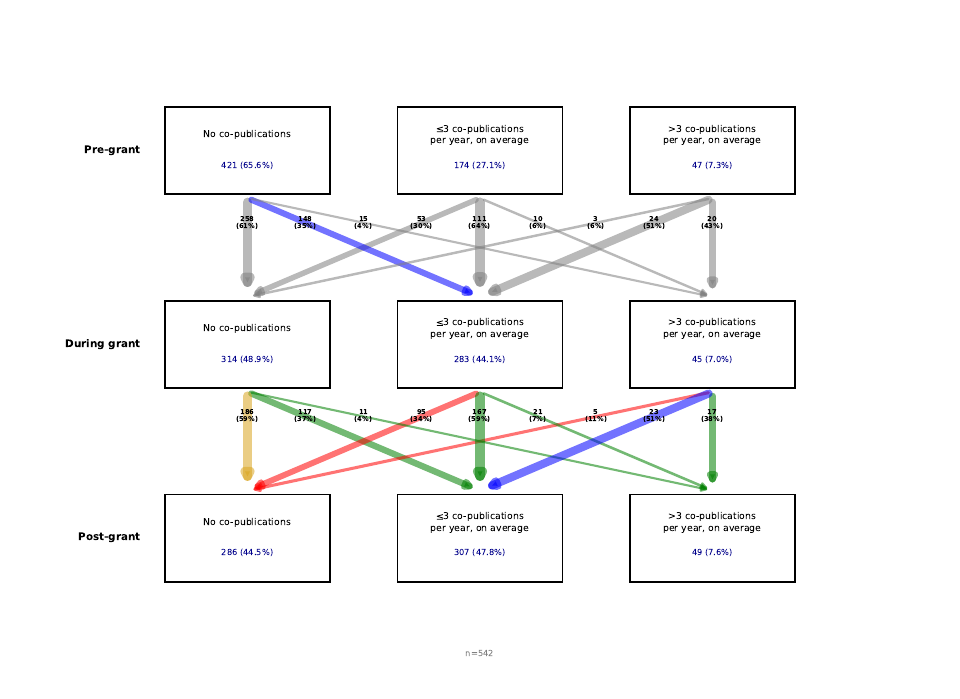}
        \caption{Fr\'{e}chet: pre=8, dur=4, post=5}
    \end{subfigure}
    \hfill
    \begin{subfigure}[t]{0.48\textwidth}
        \centering
        \includegraphics[width=\textwidth]{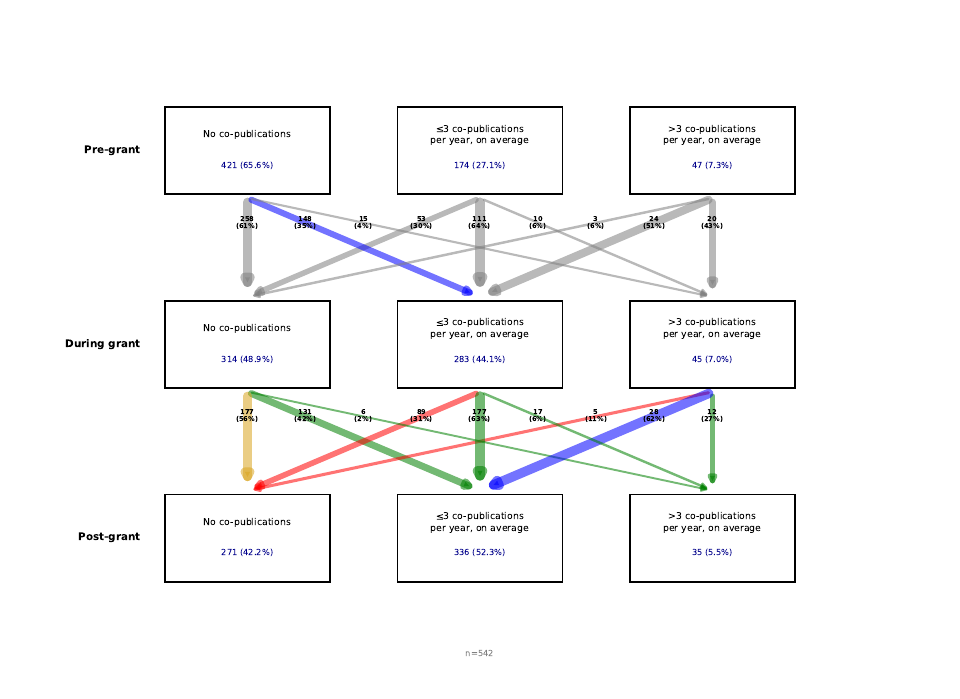}
        \caption{Fr\'{e}chet: pre=8, dur=4, post=7}
    \end{subfigure}
    \\[1ex]
    \begin{subfigure}[t]{0.48\textwidth}
        \centering
        \includegraphics[width=\textwidth]{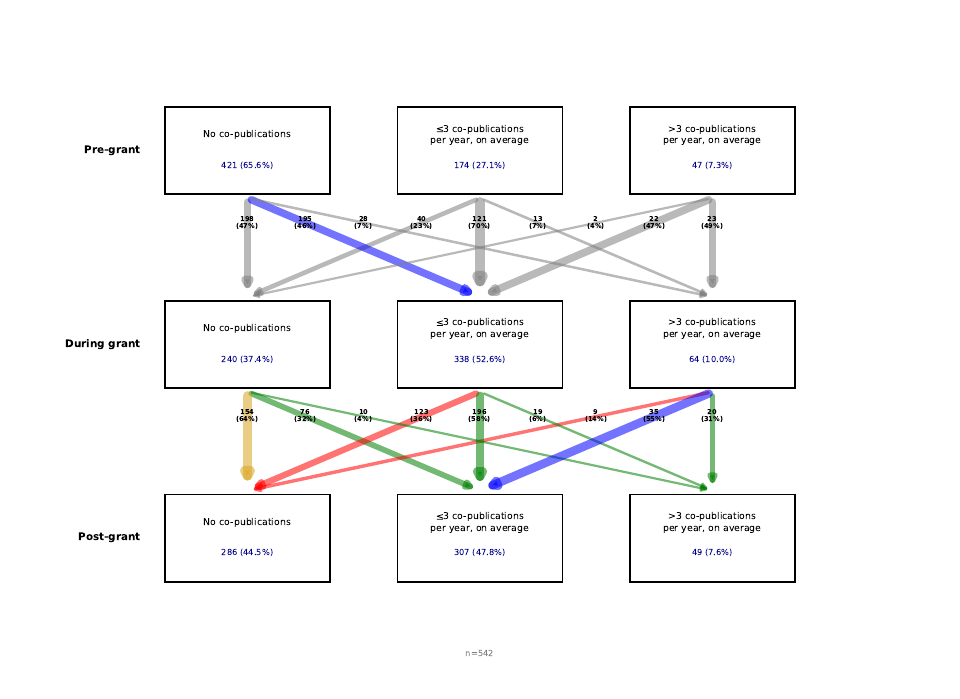}
        \caption{Fr\'{e}chet: pre=8, dur=5, post=5}
    \end{subfigure}
    \hfill
    \begin{subfigure}[t]{0.48\textwidth}
        \centering
        \includegraphics[width=\textwidth]{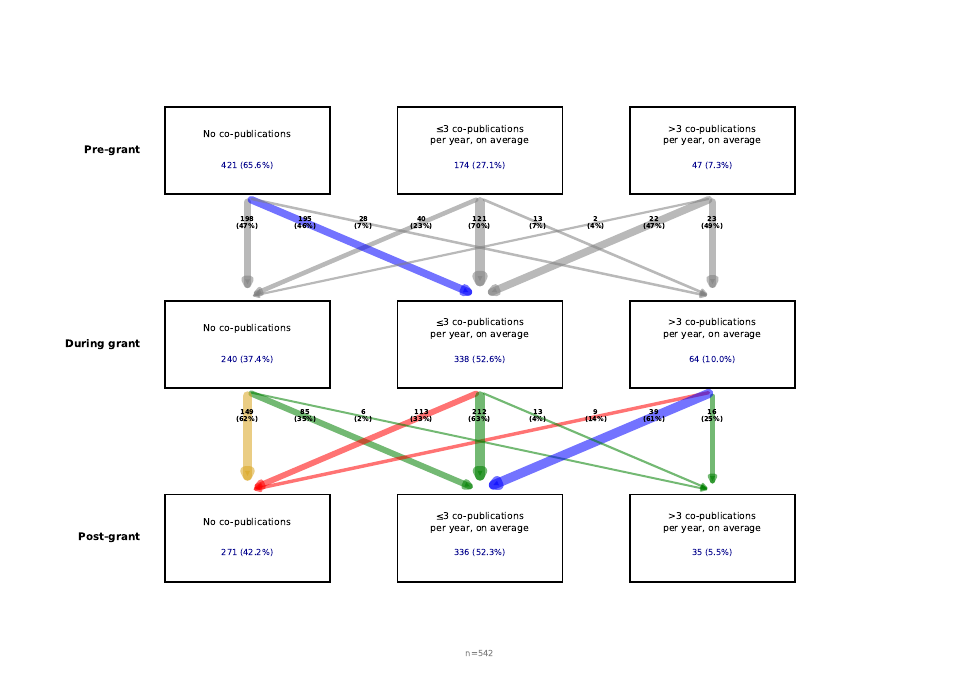}
        \caption{Fr\'{e}chet: pre=8, dur=5, post=7}
    \end{subfigure}
    \\[1ex]
    \begin{subfigure}[t]{0.48\textwidth}
        \centering
        \includegraphics[width=\textwidth]{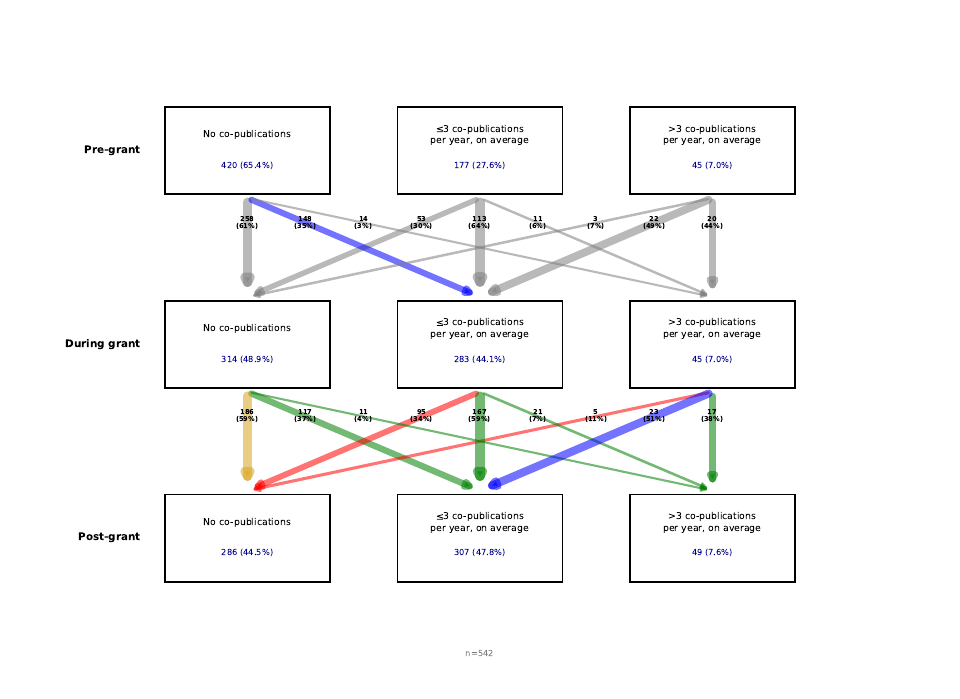}
        \caption{Fr\'{e}chet: pre=9, dur=4, post=5}
    \end{subfigure}
    \hfill
    \begin{subfigure}[t]{0.48\textwidth}
        \centering
        \includegraphics[width=\textwidth]{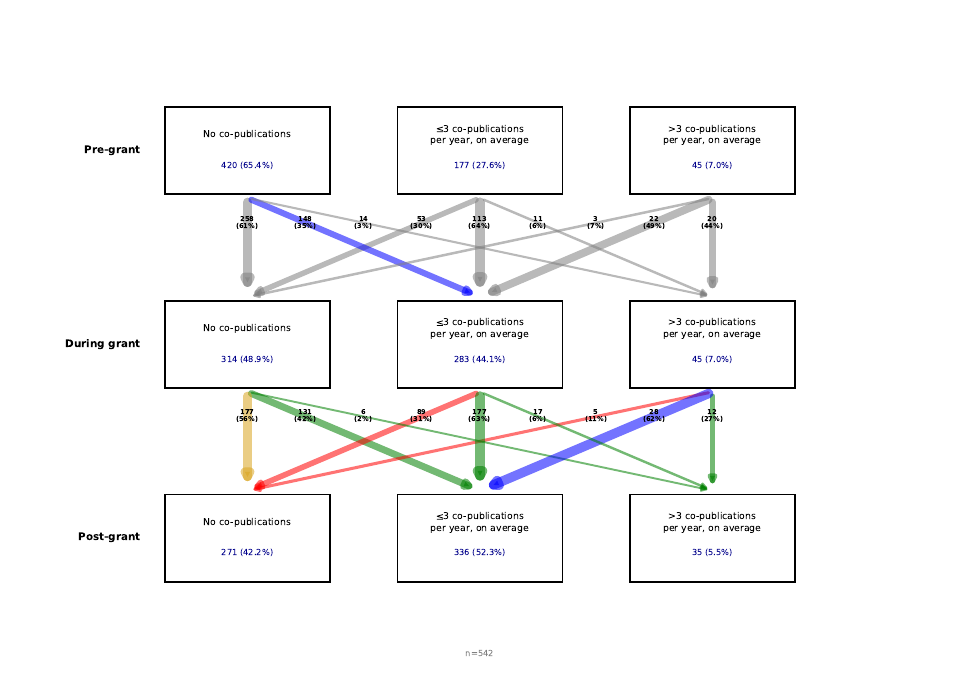}
        \caption{Fr\'{e}chet: pre=9, dur=4, post=7}
    \end{subfigure}
    \caption{Robustness analysis: cluster-flow diagrams
    (part 1 of 3). Arrow colours follow the same convention as
    Figure~\ref{fig:transitions}.}
    \label{fig:robustness_panel}
\end{figure}

\begin{figure}[H]\ContinuedFloat
    \centering
    \begin{subfigure}[t]{0.48\textwidth}
        \centering
        \includegraphics[width=\textwidth]{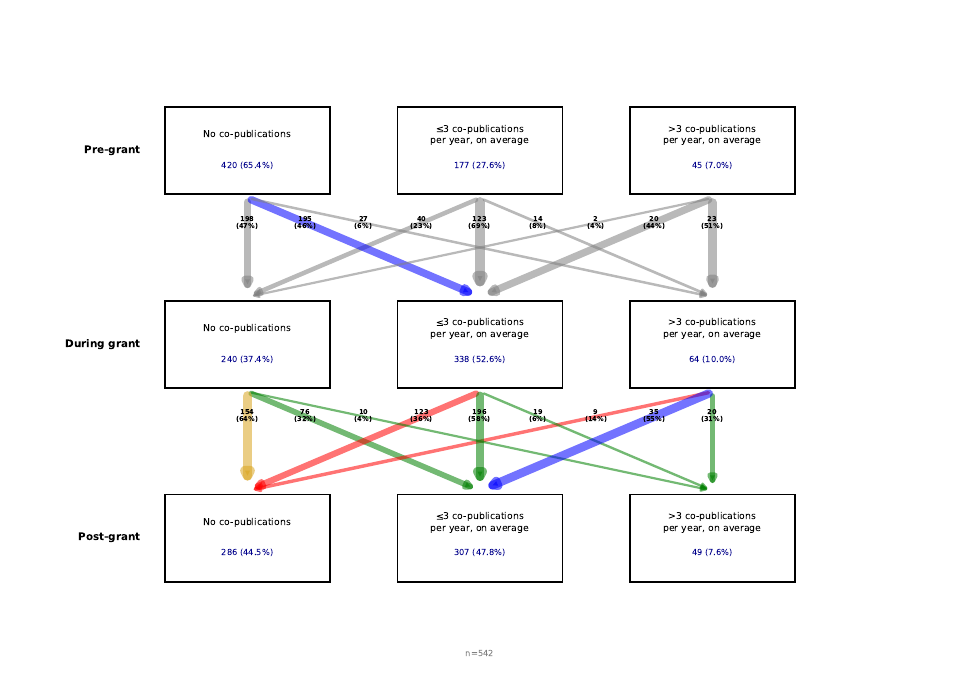}
        \caption{Fr\'{e}chet: pre=9, dur=5, post=5}
    \end{subfigure}
    \hfill
    \begin{subfigure}[t]{0.48\textwidth}
        \centering
        \includegraphics[width=\textwidth]{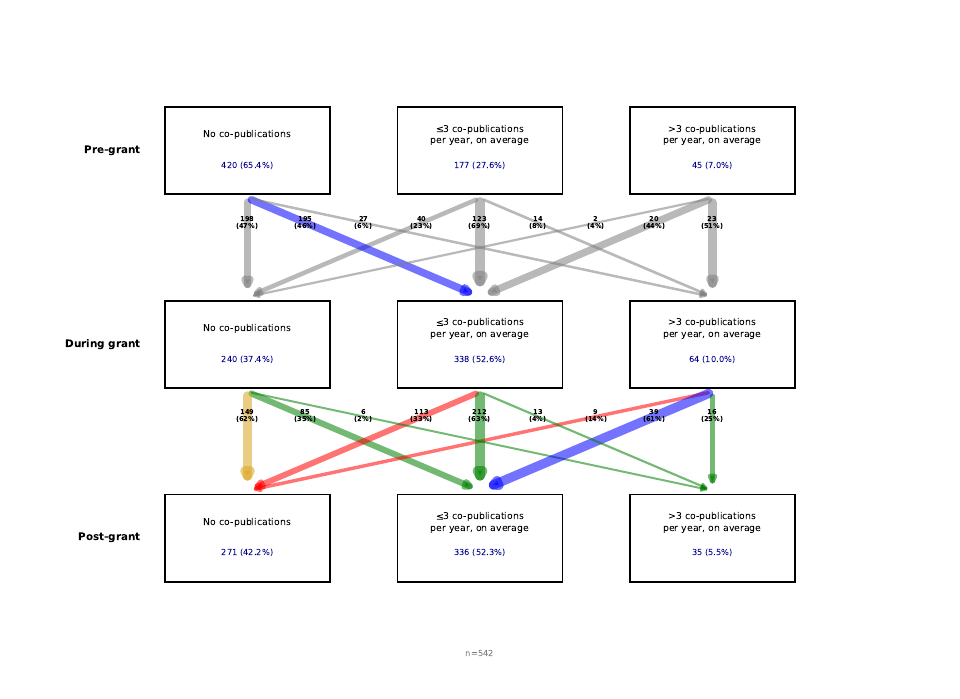}
        \caption{Fr\'{e}chet: pre=9, dur=5, post=7}
    \end{subfigure}
    \\[1ex]
    \begin{subfigure}[t]{0.48\textwidth}
        \centering
        \includegraphics[width=\textwidth]{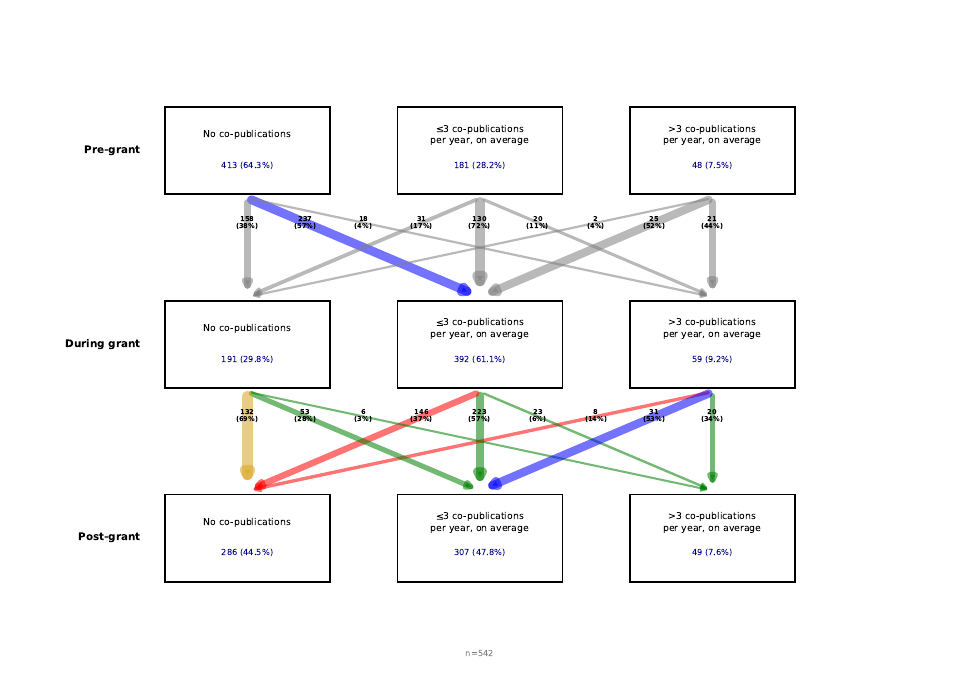}
        \caption{Fr\'{e}chet: pre=10, dur=6, post=5 (baseline windows)}
    \end{subfigure}
    \hfill
    \begin{subfigure}[t]{0.48\textwidth}
        \centering
        \includegraphics[width=\textwidth]{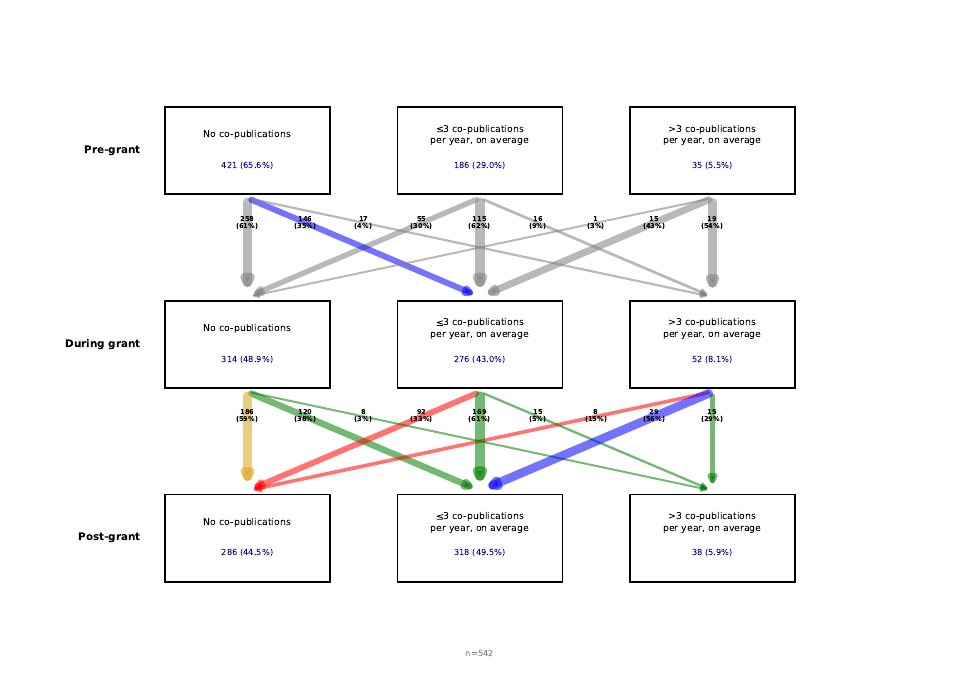}
        \caption{DTW: pre=8, dur=4, post=5}
    \end{subfigure}
    \\[1ex]
    \begin{subfigure}[t]{0.48\textwidth}
        \centering
        \includegraphics[width=\textwidth]{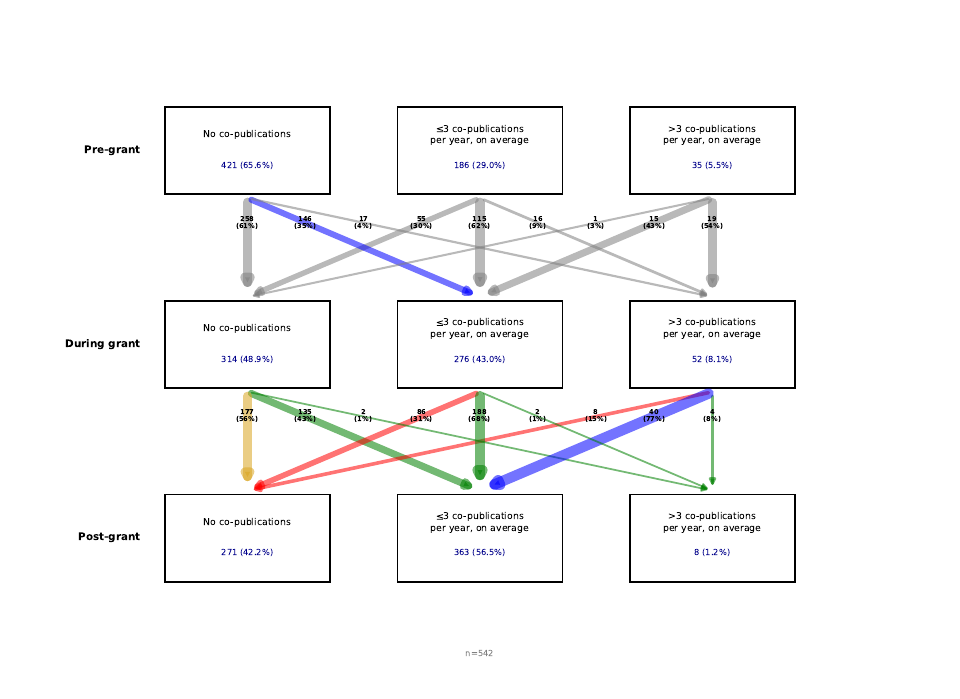}
        \caption{DTW: pre=8, dur=4, post=7}
    \end{subfigure}
    \hfill
    \begin{subfigure}[t]{0.48\textwidth}
        \centering
        \includegraphics[width=\textwidth]{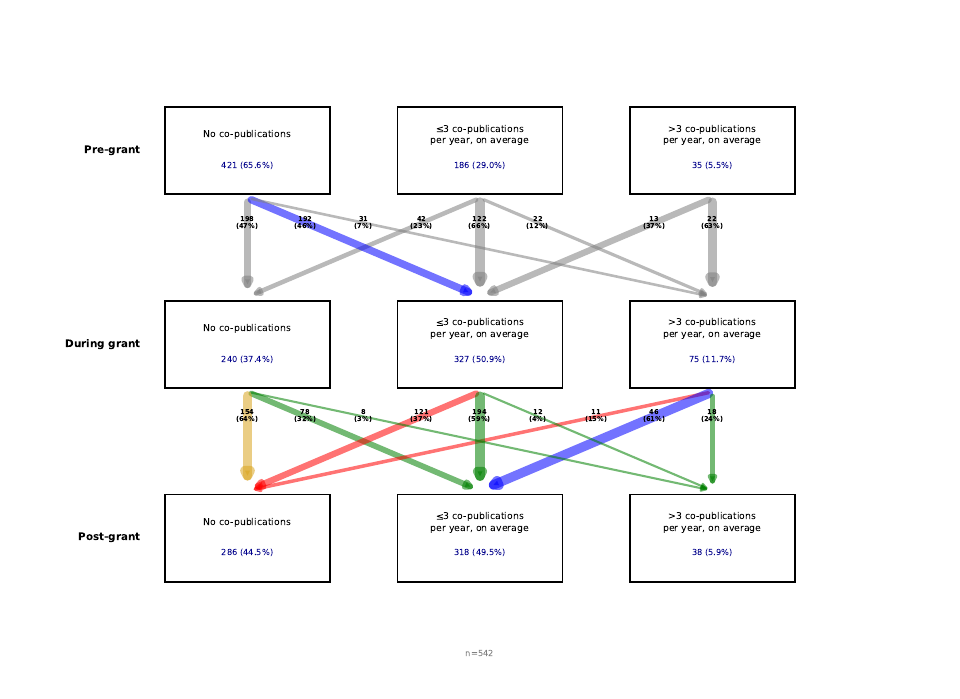}
        \caption{DTW: pre=8, dur=5, post=5}
    \end{subfigure}
    \caption{Robustness analysis: cluster-flow diagrams
    (part 2 of 3).}
\end{figure}

\begin{figure}[H]\ContinuedFloat
    \centering
    \begin{subfigure}[t]{0.48\textwidth}
        \centering
        \includegraphics[width=\textwidth]{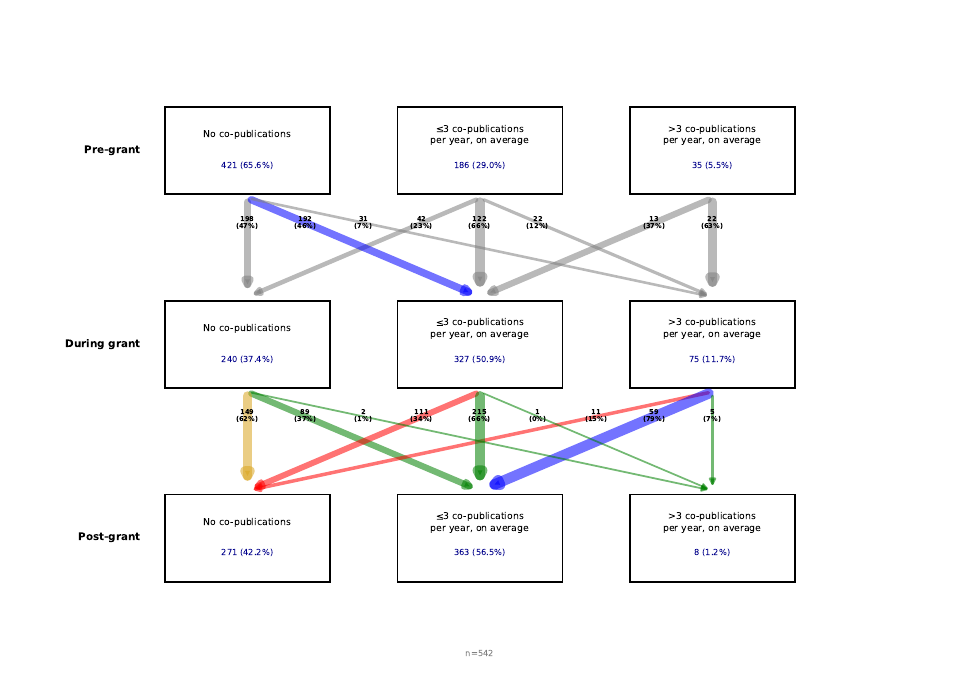}
        \caption{DTW: pre=8, dur=5, post=7}
    \end{subfigure}
    \hfill
    \begin{subfigure}[t]{0.48\textwidth}
        \centering
        \includegraphics[width=\textwidth]{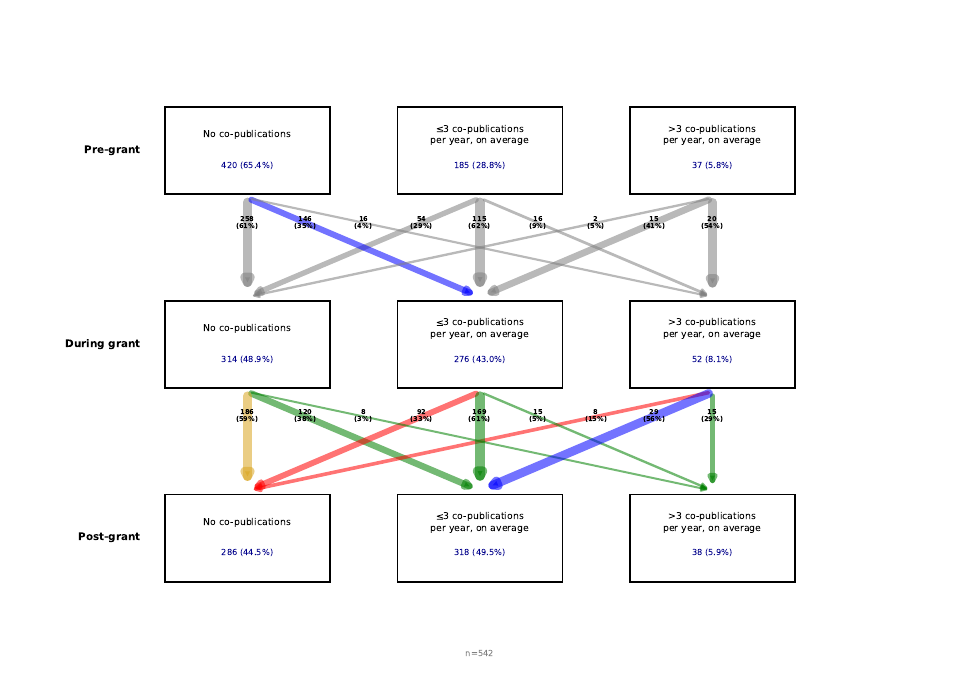}
        \caption{DTW: pre=9, dur=4, post=5}
    \end{subfigure}
    \\[1ex]
    \begin{subfigure}[t]{0.48\textwidth}
        \centering
        \includegraphics[width=\textwidth]{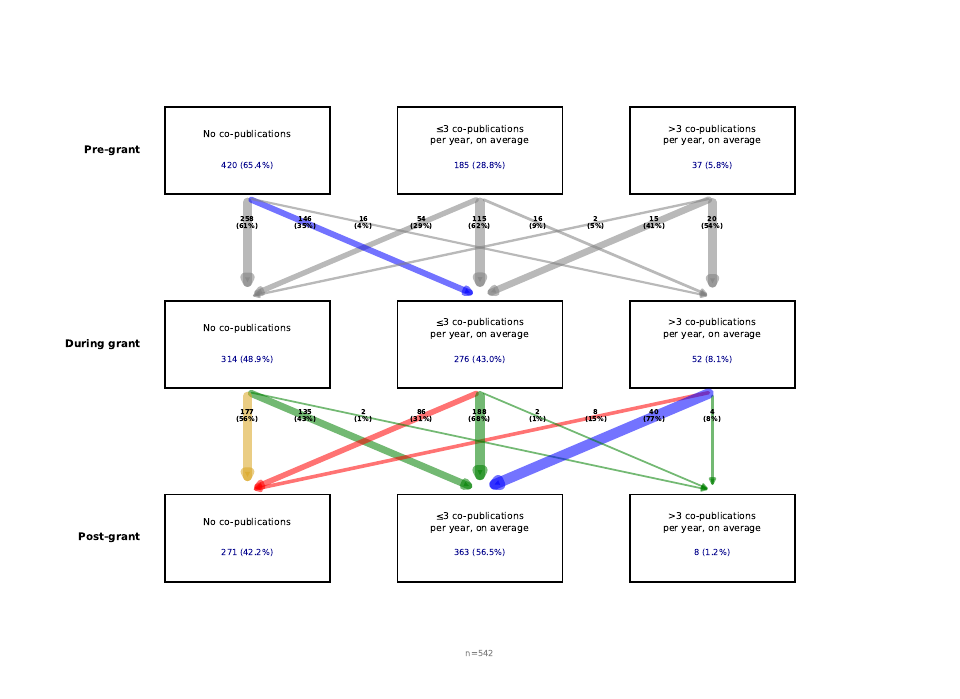}
        \caption{DTW: pre=9, dur=4, post=7}
    \end{subfigure}
    \hfill
    \begin{subfigure}[t]{0.48\textwidth}
        \centering
        \includegraphics[width=\textwidth]{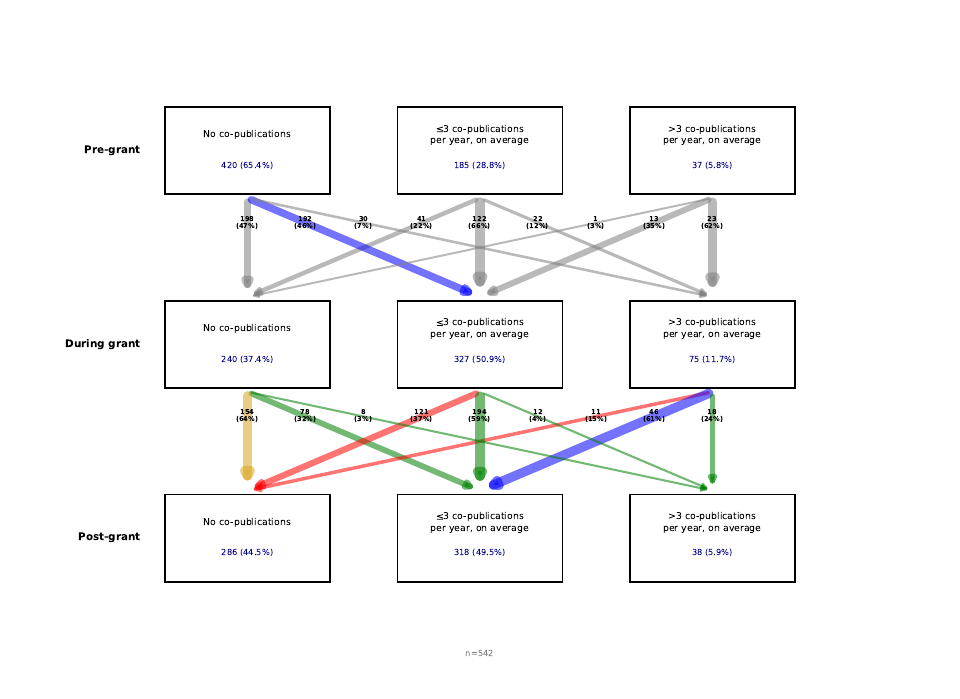}
        \caption{DTW: pre=9, dur=5, post=5}
    \end{subfigure}
    \\[1ex]
    \begin{subfigure}[t]{0.48\textwidth}
        \centering
        \includegraphics[width=\textwidth]{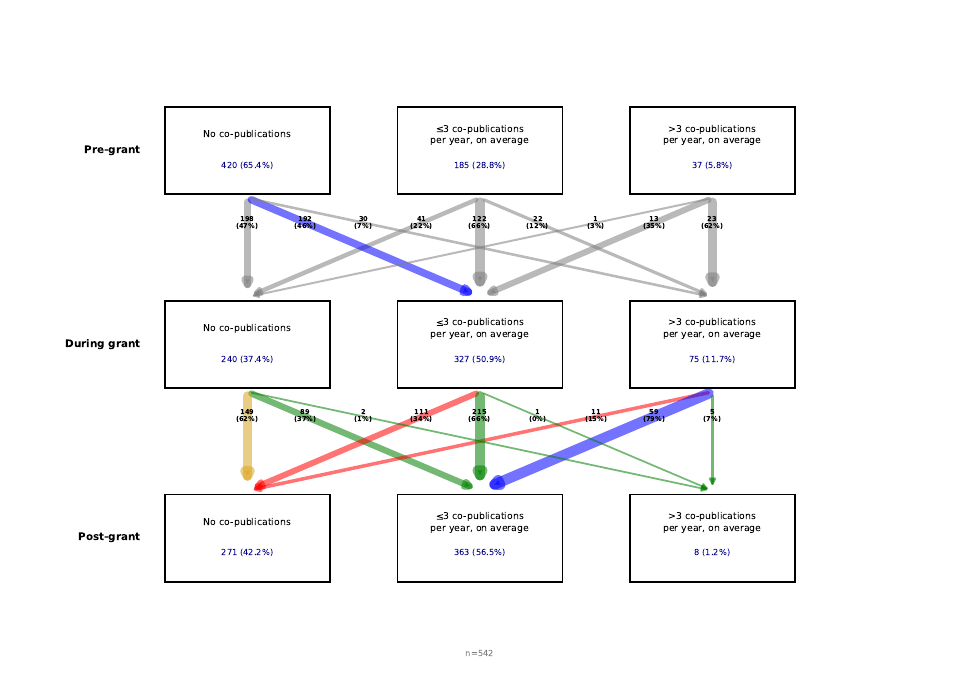}
        \caption{DTW: pre=9, dur=5, post=7}
    \end{subfigure}
    \caption{Robustness analysis: cluster-flow diagrams
    (part 3 of 3). All 17 configurations consistently recover the
    same three-cluster structure and transition patterns reported in
    the main text.}
\end{figure}

\subsection*{Feature analysis}
\label{app:feature_analysis}
\begin{figure}[ht]
    \centering
    \begin{subfigure}[t]{0.49\textwidth}
        \centering
        \includegraphics[width=\textwidth]{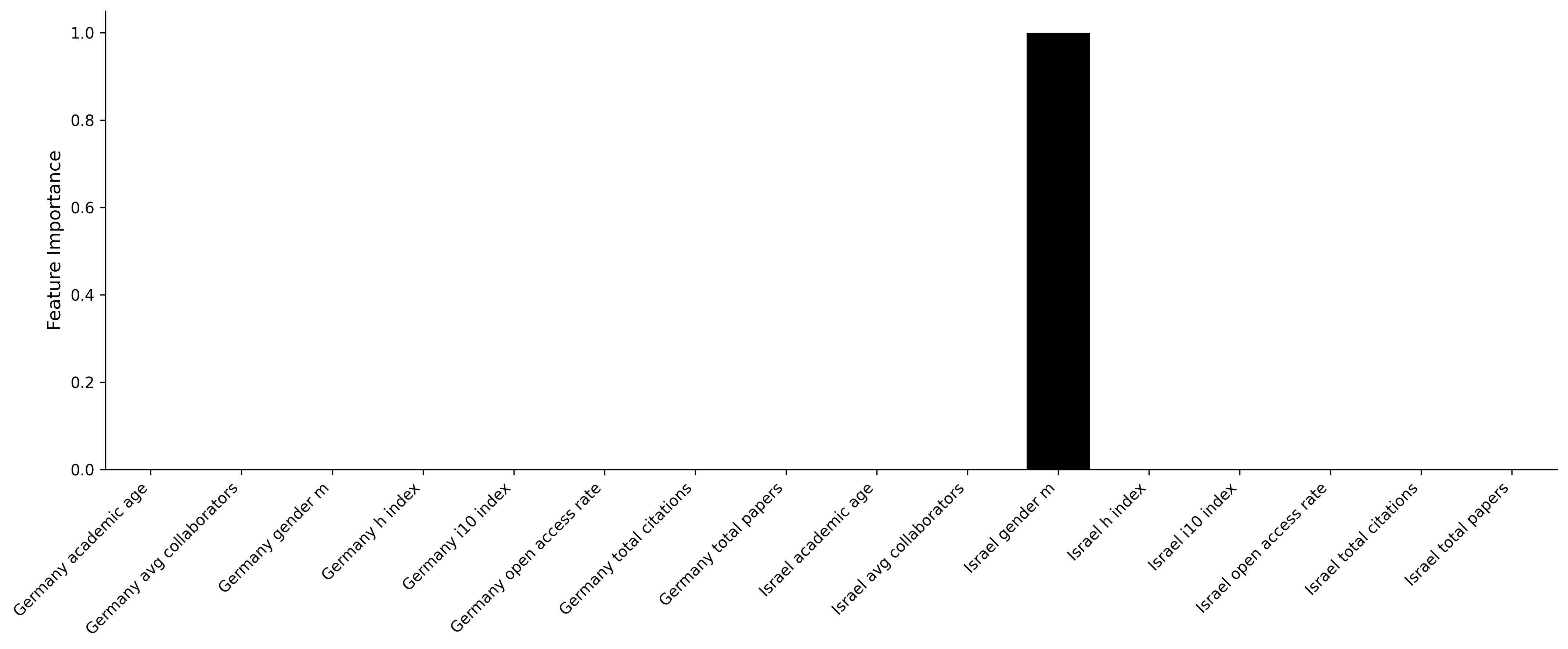}
        \caption{Logistic Regression:
        Pre $\rightarrow$ During.}
    \end{subfigure}%
    \hfill
    \begin{subfigure}[t]{0.49\textwidth}
        \centering
        \includegraphics[width=\textwidth]{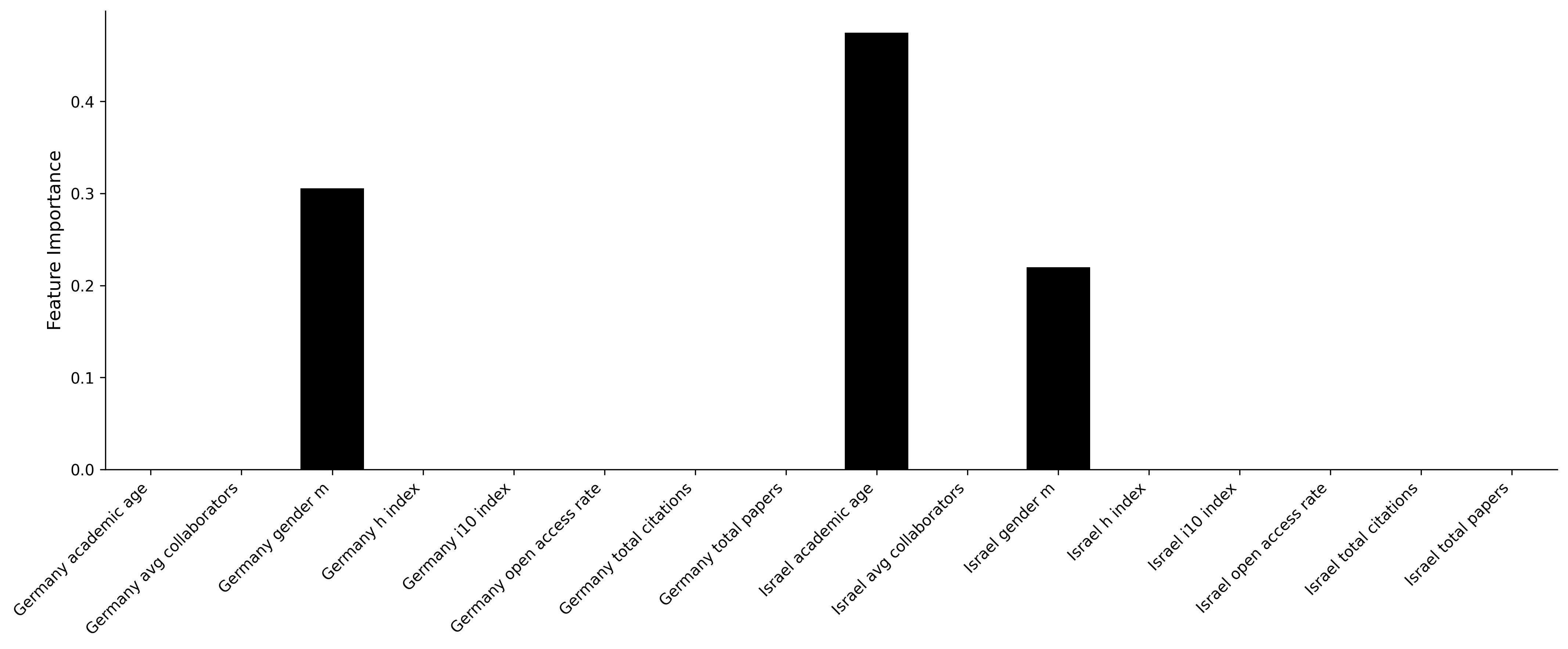}
        \caption{Logistic Regression:
        During $\rightarrow$ Post.}
    \end{subfigure}
    \\[1ex]
    \begin{subfigure}[t]{0.49\textwidth}
        \centering
        \includegraphics[width=\textwidth]{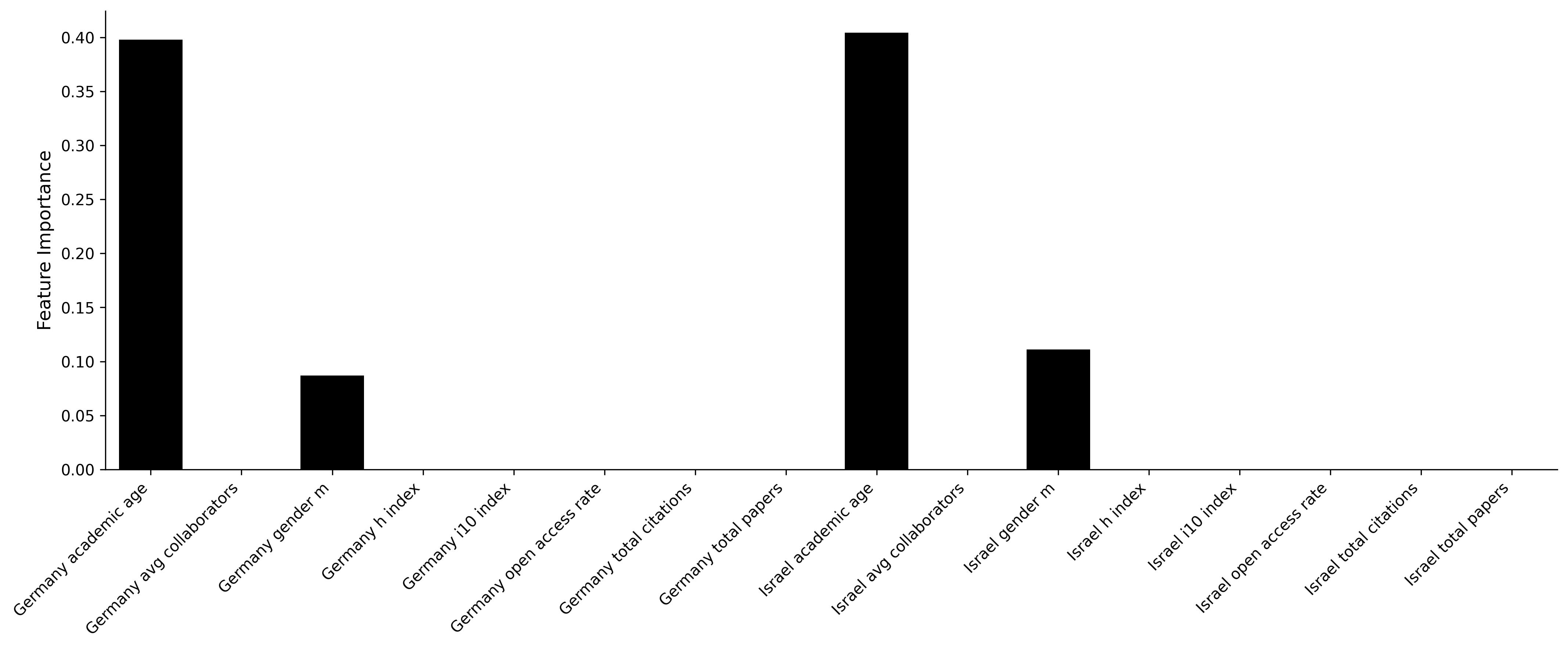}
        \caption{Random Forest: Pre $\rightarrow$ During.}
    \end{subfigure}%
    \hfill
    \begin{subfigure}[t]{0.49\textwidth}
        \centering
        \includegraphics[width=\textwidth]{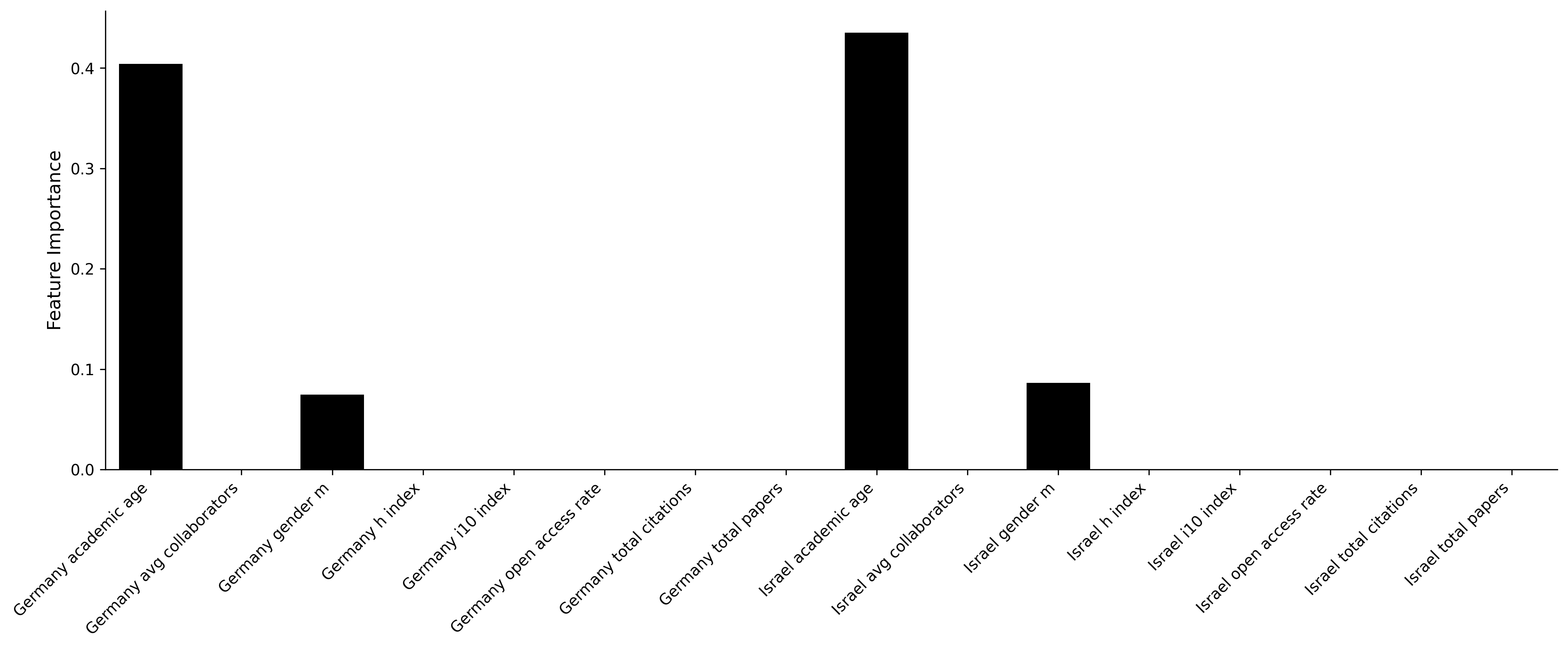}
        \caption{Random Forest: During $\rightarrow$ Post.}
    \end{subfigure}
    \\[1ex]
    \begin{subfigure}[t]{0.49\textwidth}
        \centering
        \includegraphics[width=\textwidth]{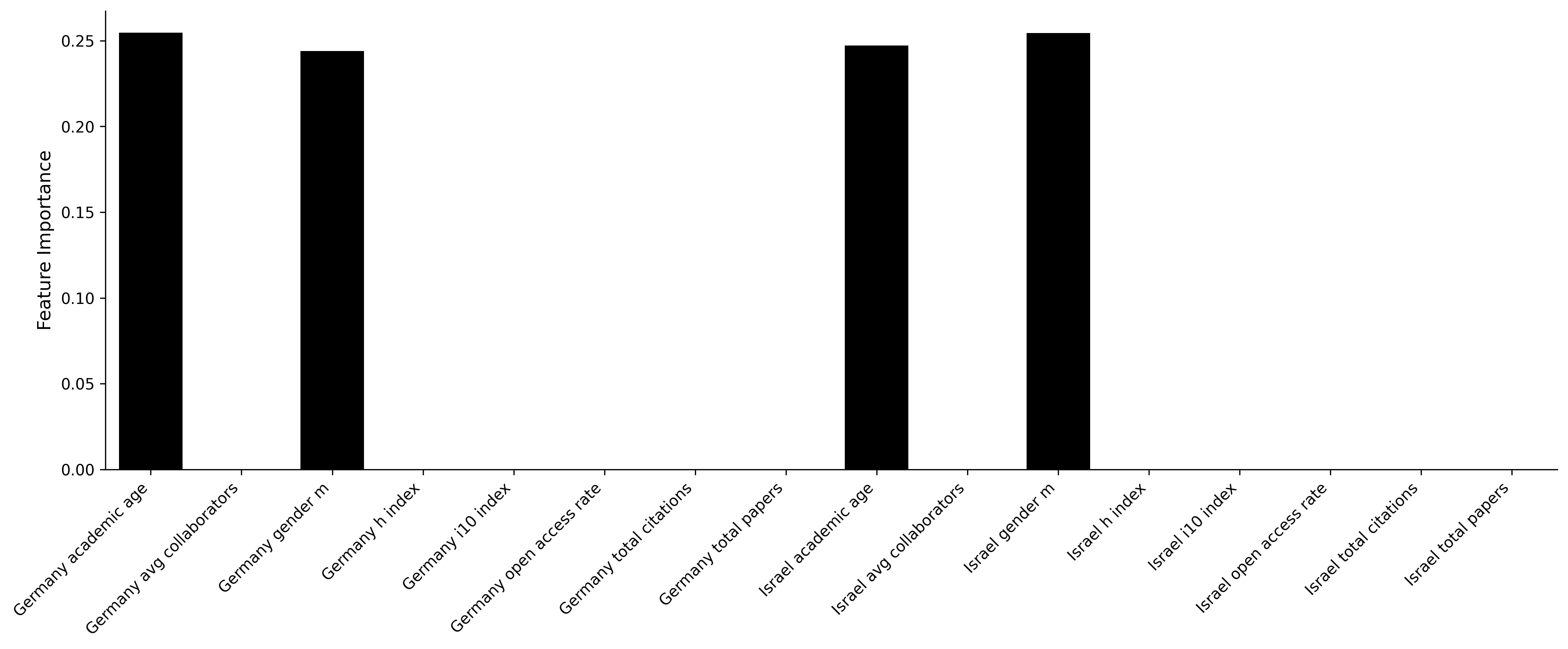}
        \caption{XGBoost: Pre $\rightarrow$ During.}
    \end{subfigure}%
    \hfill
    \begin{subfigure}[t]{0.49\textwidth}
        \centering
        \includegraphics[width=\textwidth]{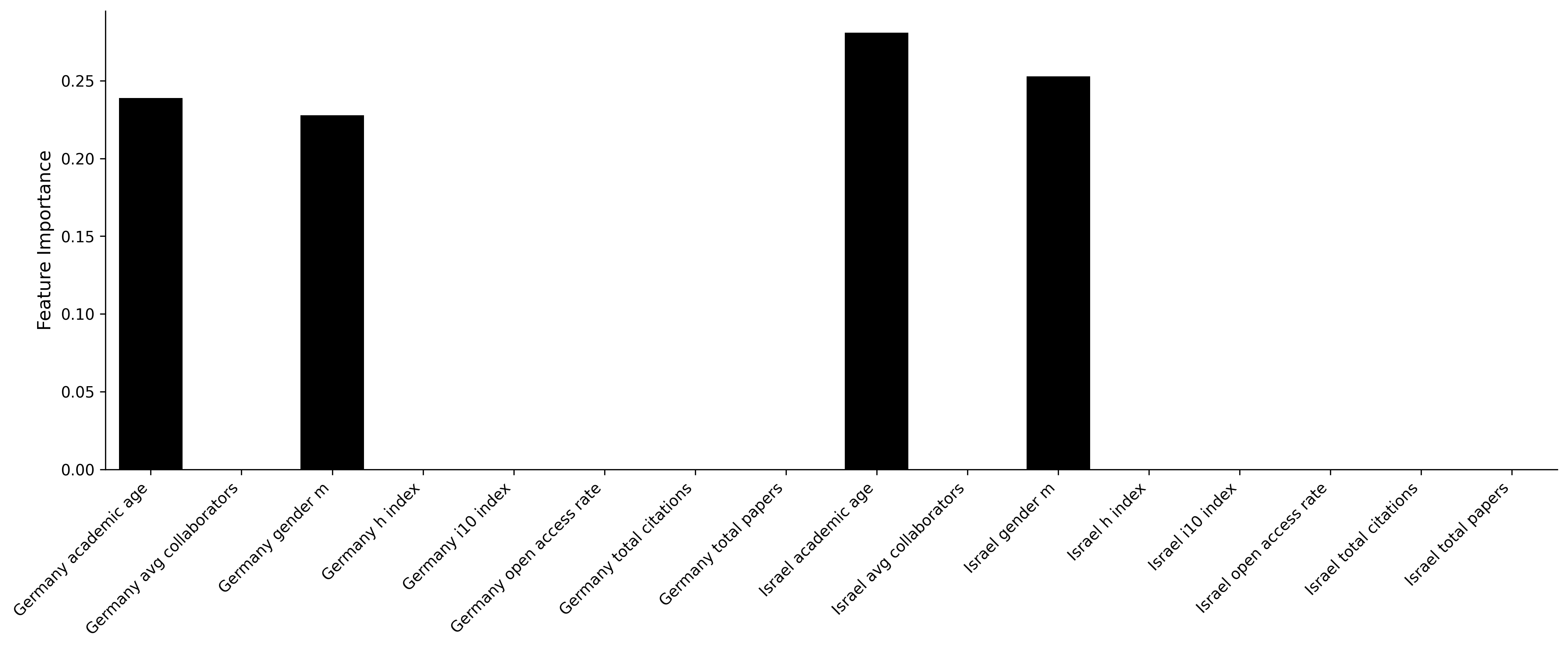}
        \caption{XGBoost: During $\rightarrow$ Post.}
    \end{subfigure}
    \caption{Feature importance distribution for each ML-based
    cluster prediction model across both transitions. Left column:
    Pre~$\rightarrow$~During; Right column:
    During~$\rightarrow$~Post.}
    \label{fig:fi}
\end{figure}

We computed feature importance for each model across both transitions
(Figure~\ref{fig:fi}). For the Pre~$\rightarrow$~During transition,
none of the three algorithms identifies features that consistently
surpass the noise threshold, reflecting the limited predictive signal
available in pre-grant bibliometrics. For the
During~$\rightarrow$~Post transition, XGBoost identifies four features
above the noise threshold: average academic age of Israeli researchers
(0.28), gender composition of the Israeli team (0.25), average
academic age of German researchers (0.24), and gender composition of
the German team (0.23). Importantly, standard impact metrics ---
h-index, i10-index, total citations, and total publications --- fall
below the noise threshold across all models and both transitions,
suggesting that the modest predictive signal that does exist is
carried primarily by demographic and seniority characteristics rather
than by measures of academic output. \review{Because no model exceeded the majority-class baseline, however, these feature-importance and SHAP patterns should be read as suggestive descriptions of the limited within-model signal rather than as established determinants of co-authorship persistence.}

\begin{figure}[ht]
    \centering
    \begin{subfigure}[t]{0.49\textwidth}
        \centering
        \includegraphics[width=\textwidth]{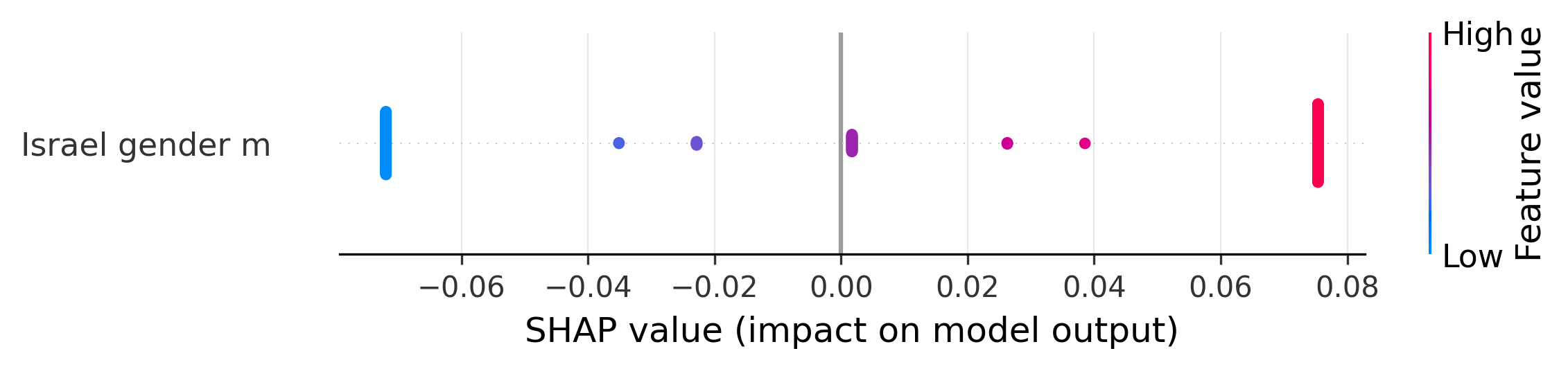}
        \caption{Logistic Regression:
        Pre $\rightarrow$ During.}
        \label{fig:shap_lg_pre}
    \end{subfigure}%
    \hfill
    \begin{subfigure}[t]{0.49\textwidth}
        \centering
        \includegraphics[width=\textwidth]{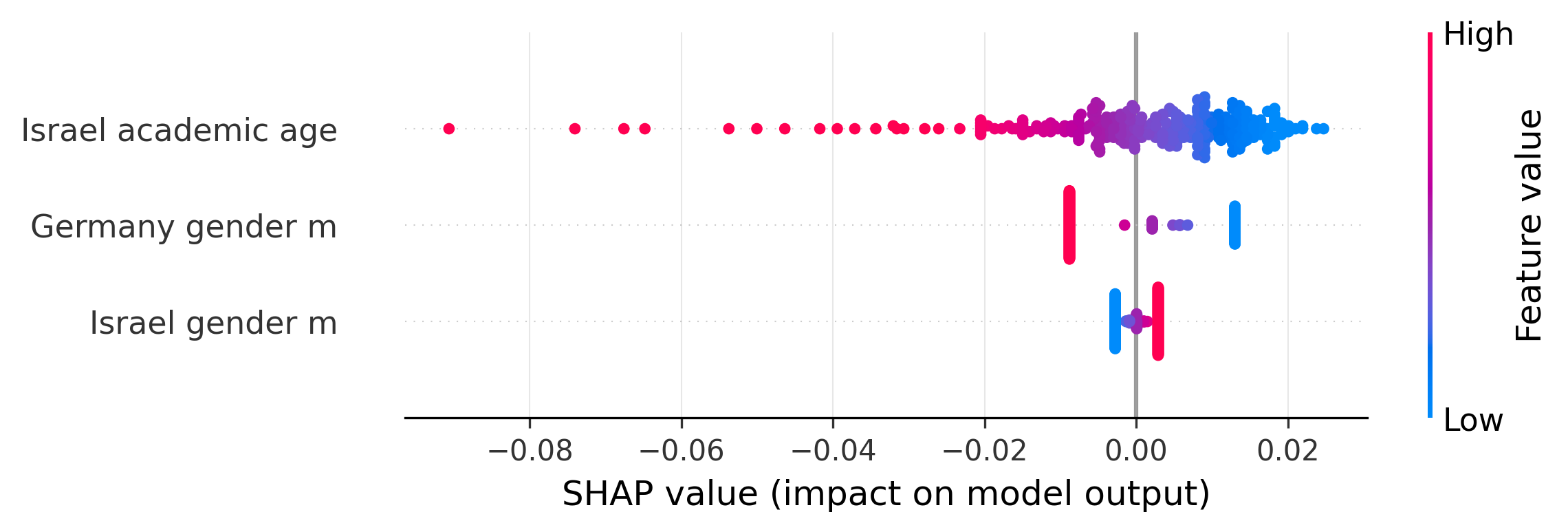}
        \caption{Logistic Regression:
        During $\rightarrow$ Post.}
        \label{fig:shap_lg_post}
    \end{subfigure}
    \\[1ex]
    \begin{subfigure}[t]{0.49\textwidth}
        \centering
        \includegraphics[width=\textwidth]{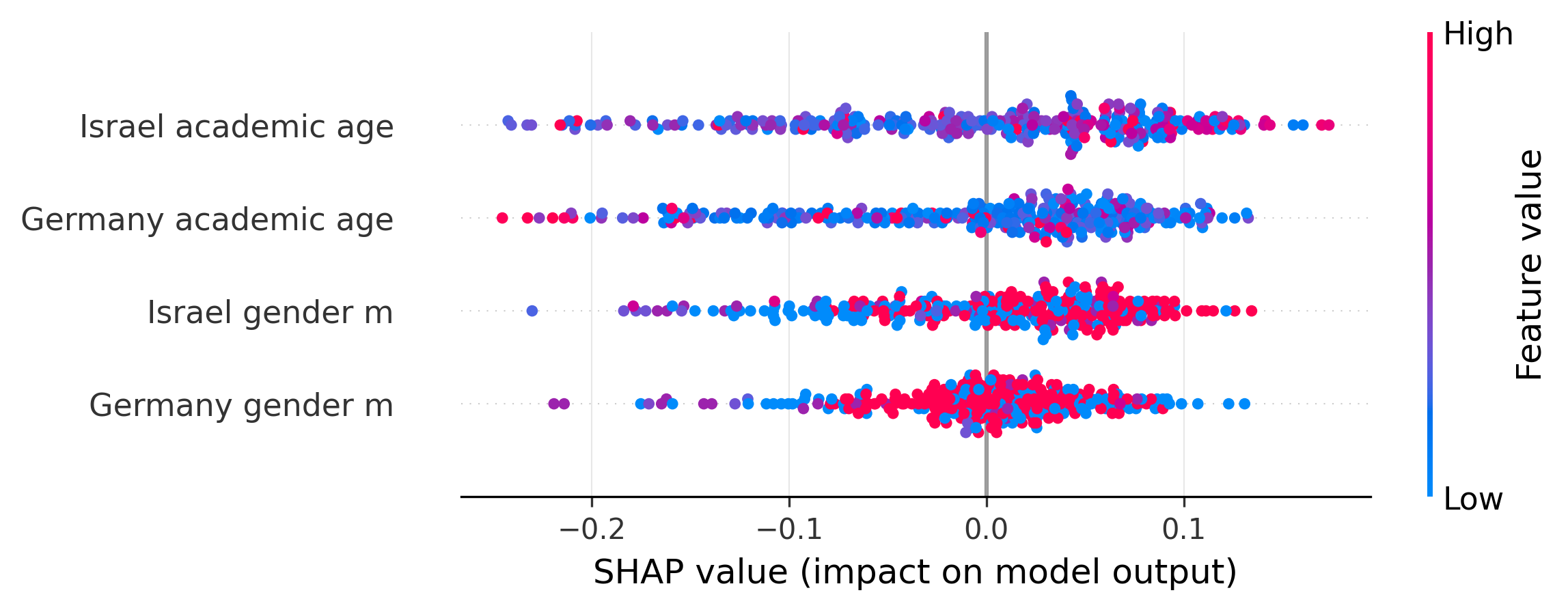}
        \caption{Random Forest:
        Pre $\rightarrow$ During.}
        \label{fig:shap_rf_pre}
    \end{subfigure}%
    \hfill
    \begin{subfigure}[t]{0.49\textwidth}
        \centering
        \includegraphics[width=\textwidth]{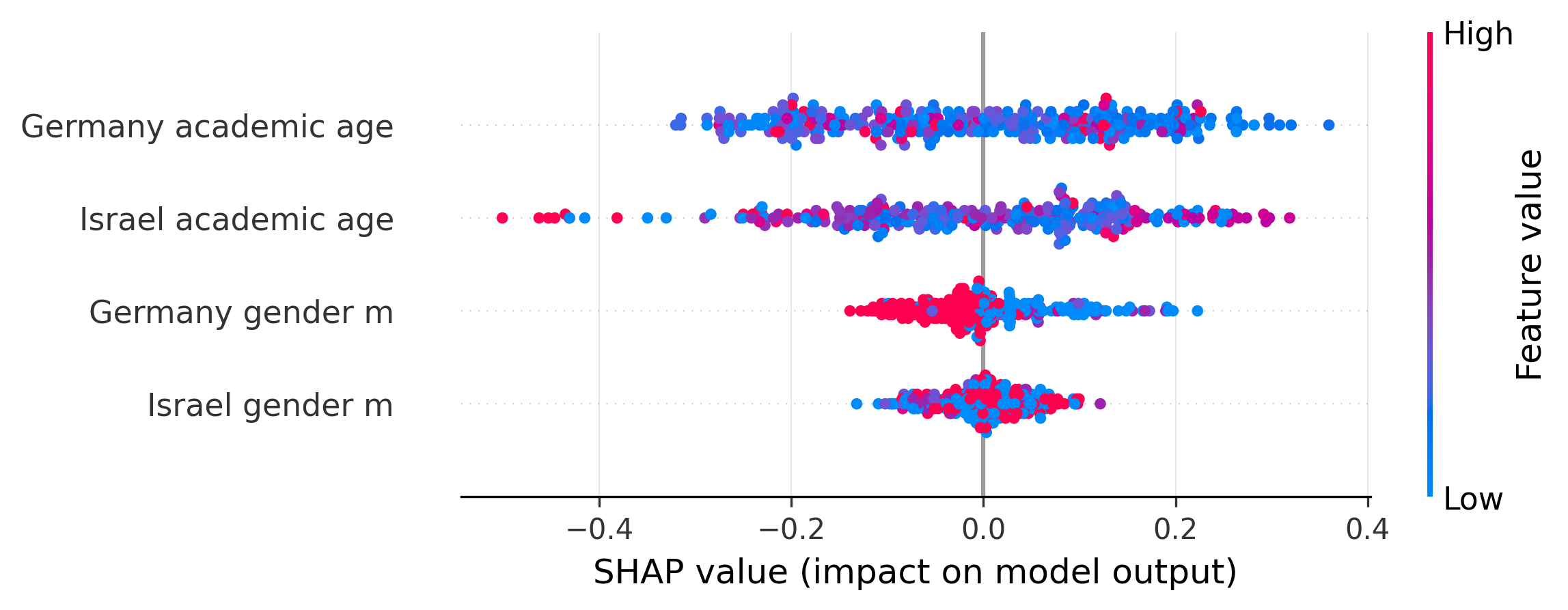}
        \caption{Random Forest:
        During $\rightarrow$ Post.}
        \label{fig:shap_rf_post}
    \end{subfigure}
    \\[1ex]
    \begin{subfigure}[t]{0.49\textwidth}
        \centering
        \includegraphics[width=\textwidth]{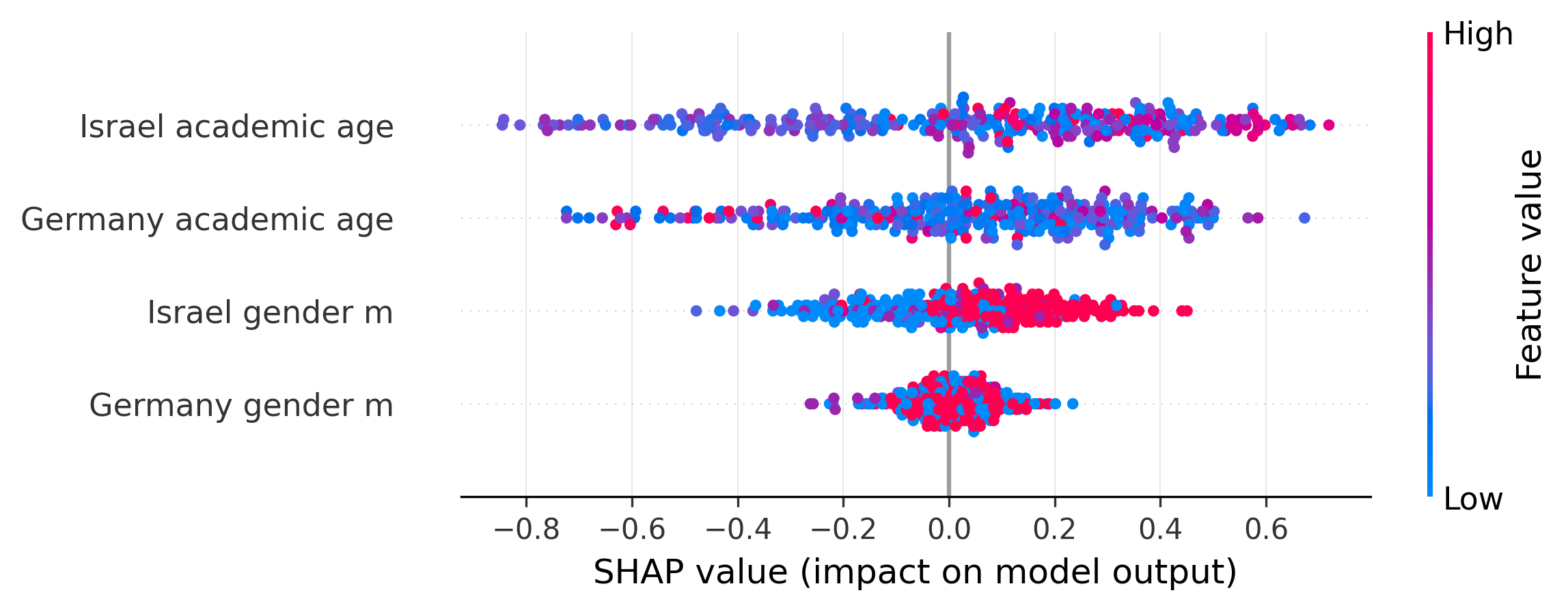}
        \caption{XGBoost: Pre $\rightarrow$ During.}
        \label{fig:shap_xgb_pre}
    \end{subfigure}%
    \hfill
    \begin{subfigure}[t]{0.49\textwidth}
        \centering
        \includegraphics[width=\textwidth]{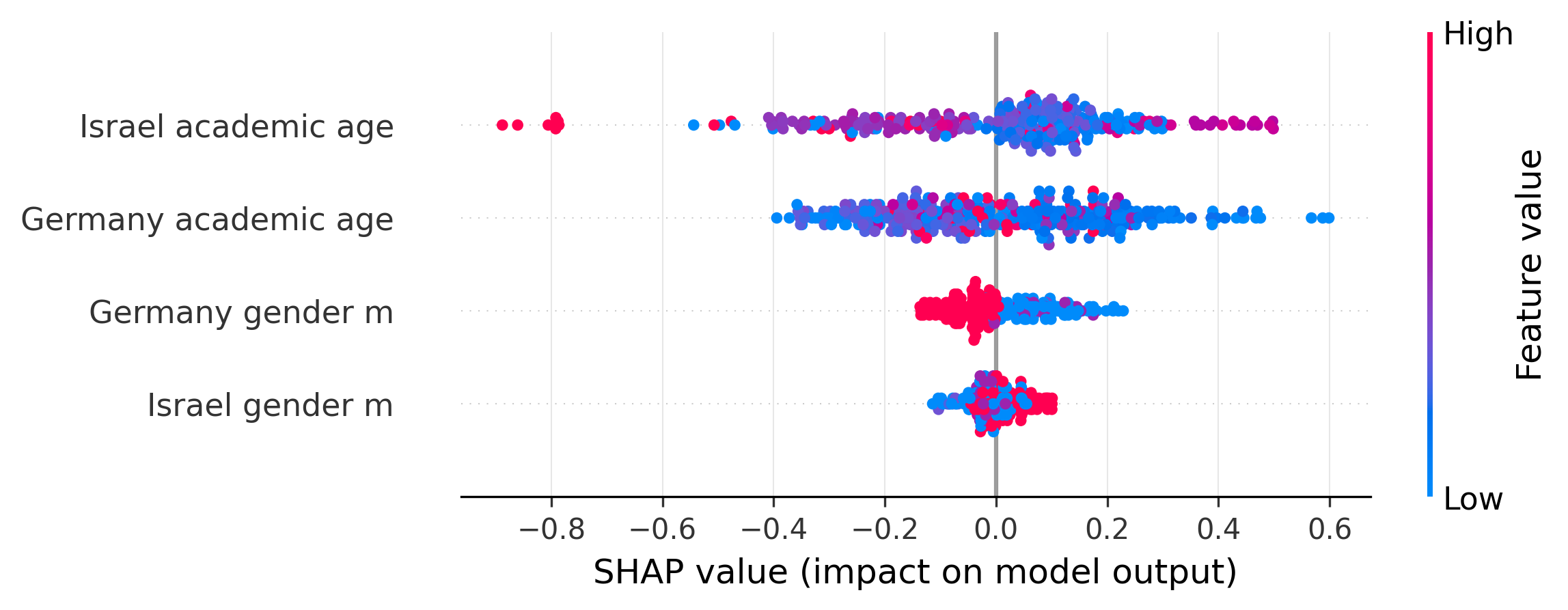}
        \caption{XGBoost: During $\rightarrow$ Post.}
        \label{fig:shap_xgb_post}
    \end{subfigure}
    \caption{SHAP analysis for each ML-based cluster
    prediction model across both transitions. Left column:
    Pre~$\rightarrow$~During; Right column:
    During~$\rightarrow$~Post.}
    \label{fig:shap}
\end{figure}

In a complementary manner, Figure~\ref{fig:shap} summarises the SHAP
(SHapley Additive exPlanations) values associated with each of the
three ML-based models across both transitions, offering a
more fine-grained perspective on how individual features influence
predictions of cluster membership. For the
Pre~$\rightarrow$~During transition (left column), no feature
consistently dominates across models, reflecting the limited
predictive signal available in pre-grant bibliometrics. For the
During~$\rightarrow$~Post transition (right column), Logistic
Regression assigns weight primarily to academic age and gender
composition, though the spread of SHAP values indicates substantial
variation across individuals. In the Random Forest model, no single
feature dominates, with all features contributing negligibly. XGBoost
provides a more balanced distribution, with academic age and gender
composition of researchers from both countries contributing
meaningfully to predictions. The SHAP distributions confirm that
higher academic age tends to push predictions toward the active
co-authorship clusters. However, the absence of strong contributions
from bibliometric impact indicators across all models and both
transitions reinforces the conclusion that co-authorship trajectory
is not reliably predicted by the scientific standing of participating
researchers.

\subsection*{Supplementary Analysis: Open-Access Publishing
Patterns among GIF Researchers}
\label{app:Open-Access}
The following analysis is offered as supplementary descriptive context,
peripheral to the central findings on co-authorship trajectories.
Publicly funded research schemes such as GIF carry an implicit
expectation that their outputs be openly accessible to the broadest
possible audience. Yet as the analysis below shows, this expectation
is far from uniformly met in practice.

While out of the scope of the main claims in this manuscript, we
include an examination of open access (OA) publishing patterns in the
context of GIF grants. Research funding, especially when it originates
from public sources, carries the expectation that funded results
circulate as widely as possible. OA has emerged as one of the key
vehicles for realising this ambition. Germany and Israel each bring
distinct traditions and infrastructures into the shared GIF framework
\cite{yair2023emotional,yair2020hierarchy,yair2019culture}.

\begin{figure}[H]
    \centering
    \includegraphics[width=1\textwidth]{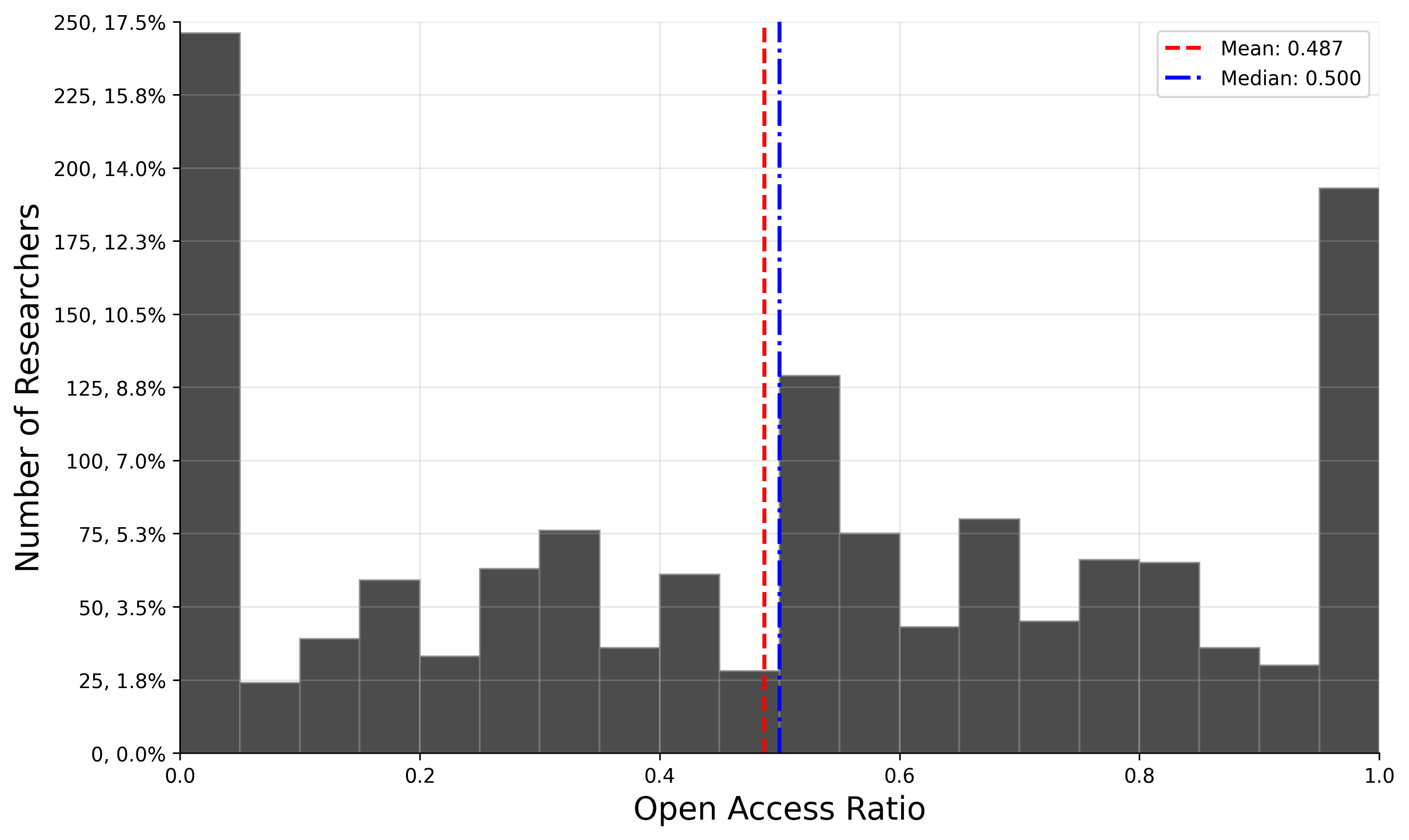}
    \caption{The distribution of grant-specific open access
    publication ratios among researchers in cluster grants.}
    \label{BBB}
\end{figure}

By examining publication patterns, we can observe how collaborations
under the GIF engage with wider debates about access and dissemination.
Figure~\ref{BBB} illustrates these dynamics: the distribution of
grant-specific OA ratios is sharply bimodal, with peaks at both
extremes: 247 researchers have no open access publications (ratio =
0.0), while nearly 192 researchers publish exclusively in open access
formats (ratio = 1.0). The mean (0.487) and median (0.500) indicate
that researchers tend to adopt either traditional publishing
strategies, mixed approaches, or fully open access models.

\end{document}